\documentclass[reprint,
superscriptaddress,
showpacs,
nofootinbib, nobibnotes,
amsmath,
amssymb,
aps,
prl]{revtex4-1}

\usepackage{microtype}
\usepackage{blindtext}
\usepackage{dcolumn}
\usepackage{bm}
\usepackage{color, soul, colortbl} 
\usepackage[colorlinks=true,
linkcolor=blue,
urlcolor=blue,
citecolor=blue,
bookmarks=true,
pdfborder={0 0 0}]{hyperref}

\usepackage{xcolor}
\usepackage{mdframed} 

\usepackage{graphicx,epsfig}
\usepackage{subfigure}
\usepackage{amsmath,amsfonts}
\usepackage{xfrac}
\usepackage{enumerate}
\usepackage[overload]{empheq}
\usepackage{mathtools}
\usepackage{cases} 
\usepackage{cancel} 

\begin{document}

\title{The physics of higher-order interactions in complex systems}

\author{Federico Battiston}
\affiliation{Department of Network and Data Science, Central European University, 1100 Vienna, Austria}
\email{battistonf@ceu.edu}

\author{Enrico Amico}
\affiliation{Institute of Bioengineering/Center for Neuroprosthetics, Ecole Polytechnique Federale de Lausanne}
\affiliation{Department of Radiology and Medical Informatics, University of Geneva, Switzerland}

\author{Alain Barrat}
\affiliation{Aix Marseille Univ, Universit\'e de Toulon, CNRS, CPT, Marseille, France}
\affiliation{Tokyo Tech World Research Hub Initiative (WRHI), Tokyo Institute of Technology, Tokyo, Japan}

\author{Ginestra Bianconi}
\affiliation{School of Mathematical Sciences, Queen Mary University of London, London E1 4NS, United Kingdom}
\affiliation{The Alan Turing Institute, The British Library,6 Euston Road, London,  NW1 2DB, United Kingdom}

\author{Guilherme Ferraz de Arruda}
\affiliation{Mathematics and Complex Systems Research Area, ISI Foundation, via Chisola 5, Turin, Italy}

\author{Benedetta Franceschiello}
\affiliation{Laboratory for Investigative Neurophysiology (The LINE), Department of Radiology, Lausanne University Hospital}
\affiliation{University of Lausanne (CHUV-UNIL), Lausanne, Switzerland}

\author{Iacopo Iacopini}
\affiliation{Department of Network and Data Science, Central European University, 1100 Vienna, Austria}

\author{Sonia K\'efi}
\affiliation{ISEM, CNRS, Univ. Montpellier, IRD, EPHE, Montpellier, France}
\affiliation{Santa Fe Institute, 1399 Hyde Park Road, Santa Fe, NM, 87501 USA}

\author{Vito Latora}
\affiliation{School of Mathematical Sciences, Queen Mary University of London, London E1 4NS, United Kingdom}
\affiliation{Dipartimento di Fisica ed Astronomia, Universit\`a di Catania}
\affiliation{INFN Sezione di Catania, 95123, Catania, Italy}
\affiliation{Complexity Science Hub Vienna (CSHV), Vienna, Austria}

\author{Yamir Moreno}
\affiliation{Mathematics and Complex Systems Research Area, ISI Foundation, via Chisola 5, Turin, Italy}
\affiliation{Complexity Science Hub Vienna (CSHV), Vienna, Austria}
\affiliation{Institute for Biocomputation and Physics of Complex Systems (BIFI), University of Zaragoza, 50018 Zaragoza, Spain}

\author{Micah M. Murray}
\affiliation{Laboratory for Investigative Neurophysiology (The LINE), Department of Radiology, Lausanne University Hospital}
\affiliation{CIBM Center for Biomedical Imaging,
Lausanne, Switzerland}
\affiliation{Department of Hearing and Speech Sciences, Vanderbilt University, Nashville, TN, USA}

\author{Tiago P. Peixoto}
\affiliation{Department of Network and Data Science, Central European University, 1100 Vienna, Austria}
\affiliation{Department of Mathematical Sciences, University of Bath, Claverton Down, Bath BA2 7AY, United Kingdom}

\author{Francesco Vaccarino}
\affiliation{Dipartimento di Scienze Matematiche, Politecnico di Torino, Turin, Italy \&
SmartData@PoliTO, Politecnico di Torino, Turin, Italy}

\author{Giovanni Petri}
\affiliation{Mathematics and Complex Systems Research Area, ISI Foundation, via Chisola 5, Turin, Italy}
\affiliation{ISI Global Science Foundation, 33 W 42nd St, 10036 New York NY, USA}
\email{giovanni.petri@isi.it}

\begin{abstract}
Complex networks have become the main paradigm for modelling the dynamics of interacting systems. However, networks are intrinsically limited to describing pairwise interactions, whereas real-world systems are often characterized by higher-order interactions involving groups of three or more units. Higher-order structures, such as hypergraphs and simplicial complexes, are therefore a better tool to map the real organization of many social, biological and man-made systems. Here, we highlight recent evidence of collective behaviours induced by higher-order interactions, and we outline three key challenges for the physics of higher-order systems.
\end{abstract}	
 
\maketitle

Network science helps us to better understand the evolution of the highly interconnected world in which we live~\cite{barabasi2011network}. It sheds light on myriad systems --- everything from how rumours spread in a social network to how large ecosystems stabilize in spite of competing interactions between species. A key feature shared by such systems is that they are characterized by a complex set of interactions that govern their emergent dynamics~\cite{boccaletti2006complex, barrat2008dynamical, vespignani2012modelling}. In the recent years, the architecture of social networks, ecosystems and the human brain have all been modelled as graphs, with collections of nodes describing the units of the systems --- humans, animals or neurons --- and edges encoding their pairwise interactions. This approach has led to the discovery that a heavy-tailed distribution in the number of contacts within a population causes the epidemic threshold to vanish, putting everyone at risk during a pandemic \cite{pastor2001epidemic,boguna2003absence}. It has inspired the realizations that small-world networks and clustering promote synchronization \cite{barahona2002synchronization}, and that efficient communication structures tend to reach rapid and diffused consensus, but are also prone to the spreading of misinformation \cite{del2016spreading}.

\begin{figure*}
\centering
\includegraphics[width=0.7\textwidth]{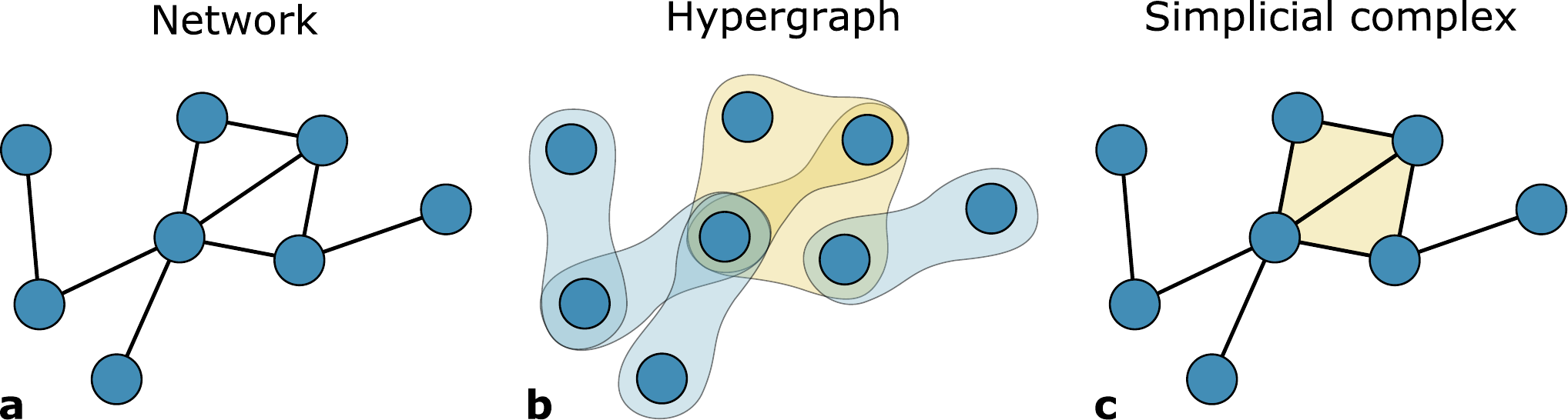}
\caption{\textbf{Pairwise and higher-order representations.}
{\bf a,} Systems comprising many interacting units have long been represented as networks, with interactions restricted to pairs of nodes and represented as edges. 
However, it is not always possible to describe group interactions as sums of pairwise interactions only. {\bf b,} Representations allowing for genuine group interactions include hypergraphs, which can encode interactions among an arbitrary number of units without further constraints. Here, shaded groups of nodes represent hyperedges. {\bf c,} Simplicial complexes offer another approach. Although more constrained than hypergraphs, they provide access to powerful mathematical formalisms \cite{patania2017topological}. Edges (1-simplices) are shown here in black, full triangles (2-simplices) in yellow. Note that in simplicial complexes, all subfaces of a simplex (for example, the edges of a triangle) need to be included. This constraint does not hold for hypergraphs.}
\label{fig:fig1}
\end{figure*}

But graphs, however convenient, can only provide a limited description of reality. They are inherently constrained to represent systems with pairwise interactions only. Yet in many biological, physical and social systems, units may interact in larger groups, and such interactions cannot always be decomposed as a linear combination of dyadic couplings~\cite{battiston2020networks} (Fig. \ref{fig:fig1}). For example, evidence from neural systems shows that higher-order effects are present and important both statistically \cite{schneidman2006weak,schneidman2003network,yu2011higher} and topologically \cite{giusti2015clique,gardner2021toroidal}. However, there is also evidence to suggest that such higher-order signatures might in some cases be redundant, and may be fully describable in terms of pairwise interactions \cite{ganmor2011sparse,merchan2016sufficiency}. 
In ecological systems, evidence clearly shows the existence of complex many-body interactions between multiple species \cite{mayfield2017higher,cervantes2020context,bairey2016high}, although the effects induced by their interaction patterns have only recently been investigated formally \cite{grilli2017higher}. 
Other examples include metabolic and genetic systems \cite{ritz2014signaling}, social coordination \cite{centola2018experimental} and group formation \cite{milojevic2014principles}.

The idea of higher-order interactions is well-known in the setting of many-body physics, for example in strong interactions \cite{povh1995many,duck1966three} or van der Waals interactions \cite{kim2006van}, as well as in statistical mechanics \cite{zeiher2017coherent}. However, in all these cases, representations of higher-order interactions are simple in the sense that they do not contribute to the emerging complexity of the problem. In complex systems typically described as networks, the story is different, and in many cases these interactions must be taken into account using more advanced mathematical structures, such as hypergraphs and simplicial complexes~\cite{battiston2020networks}. Several investigations have already shown that the presence of higher-order interactions may significantly impact the dynamics on networked systems, from diffusion~\cite{schaub2020random,carletti2020random} and synchronization~\cite{millan2019synchronization,skardal2019abrupt} to social~\cite{iacopini2019simplicial,de2020social, neuhauser2019multi} and evolutionary processes~\cite{alvarez2021evolutionary}, possibly leading to the emergence of abrupt (explosive) transitions between states. And although most research in complex systems focuses on the dynamical evolution of the states of the nodes, it is natural to consider that higher-order structures (described by hyperedges) could themselves possess a dynamical state, leading to a whole new panorama of dynamical processes. Finally, although many datasets can be easily visualized as networks, very few are readily described using a hypergraph representation. The challenge of going from the dynamics of units, and possibly information about their pairwise interactions, to a meaningful pattern of higher-order interactions between these units, remains substantial. In this Perspective, we outline the main signatures of new physics arising in higher-order systems, and we propose three key directions for future research. \\

{\bf A general pathway to explosive transitions.} 
Most processes on networks, from the dynamical evolution of coupled oscillators to the spreading of diseases, display emerging collective behaviours. Typically, such phenomena are described by continuous phase transitions: the order parameter describing, for example, the emergence of synchronization between oscillators, increases continuously as the control parameter crosses a critical threshold.
Similar transitions are also well known for percolation on networks, where 
small clusters that are initially separated  merge together to span a non-vanishing fraction of the system size at a critical point. In contrast, an explosive transition was first found some years ago for particular set of link selection rules ~\cite{achlioptas2009explosive}, for which the size of the largest cluster seemed to jump abruptly to a finite value at the transition. Although this specific transition was later classified as continuous with anomalous scaling~\cite{da2010explosive, riordan2011explosive}, explosive phenomena formed a focus of intense research activity in the years following the initial discovery~\cite{d2019explosive}. Several discontinuous phase transitions were confirmed for different processes, such as synchronization. 

\begin{figure*}
\centering
\includegraphics[width=0.8\textwidth]{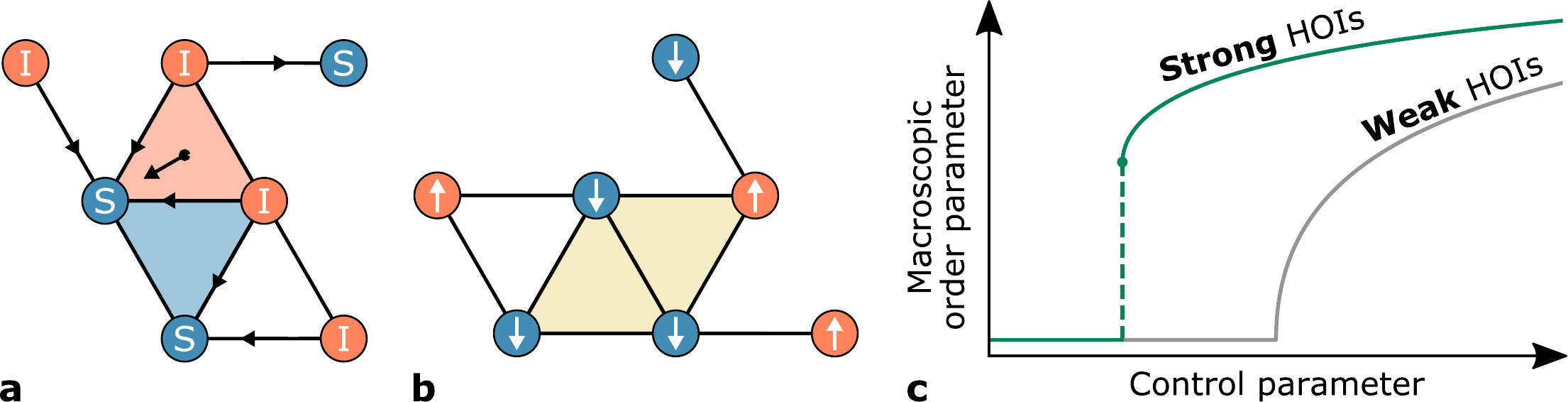}
\caption{\textbf{Higher-order interactions lead to explosive phenomena.}
Edges and hyperedges encode pairwise and group-level couplings among the nodes of a complex system. {\bf a,} Hyperedges modulate group infection and many-body feedback in higher-order processes of contagion. Susceptible nodes (S, blue) can be infected by infectious ones (I, orange) in the usual way along edges, but also by groups containing a large fraction of infected nodes (for example, orange 2-simplices). {\bf b,} Hyperedges have a similar effect on higher-order processes of synchronization, in which oscillators on nodes can be coupled along edges, or in groups via higher-order interactions. {\bf c,} Abrupt transitions emerge when increasing the strength of such interactions, suggesting a general pathway to explosive phenomena.}
\label{fig:fig2}
\end{figure*}

However, explosive phenomena are rather difficult to obtain for systems represented as networks --- those with only pairwise interactions. They can be engineered by adding artificial elements or rules to the most natural dynamical setups in an attempt to prevent the transition. Eventually though, these additions produce abrupt jumps in the order parameter once the transition becomes inevitable. For example, synchronization can become explosive in heterogeneous networks by correlating the natural frequency of oscillators to their degree~\cite{gomez2011explosive}. However, explosive phenomena are known to exist in nature, and developing a better understanding of how they behave is of key interest in many fields, primarily because
they are more difficult to 
handle, predict and control than their continuous counterparts.

A modelling approach that goes beyond networks by taking higher-order interactions into account provides a framework in which explosive phenomena emerge naturally and can therefore be studied more easily.
An abrupt transition was recently observed in a model social contagion evolving on simplicial complexes~\cite{iacopini2019simplicial}, in which individuals can assume either an infected or susceptible state. In contrast to previous proposals, here pairwise transmission does not operate alone, but can be reinforced by simplicial interactions associated with group pressure (Fig. \ref{fig:fig2}a). The model can be solved analytically with a mean-field approximation, showing that a discontinuous transition from a healthy to endemic phase (in which a significant fraction of the population is infected) emerges when the relative weight of higher-order interactions crosses a threshold. 
Interestingly, the inclusion of three-body interactions is sufficient to obtain a bistable region where endemic and non-endemic states can co-exist. 
This result has been found to be robust and general. Explosive transitions have in fact been observed in heterogeneous~\cite{matamalas2020abrupt} and time-varying~\cite{chowdhary2021simplicial, st2021bursty} structures, and in the more general setup of hypergraphs~\cite{de2020social, landry2020effect, st2021influential}, where they can also be related to higher-order discontinuous percolation processes \cite{sun2021higher}.

\begin{figure*}
	\centering
	\includegraphics[width=0.65\textwidth]{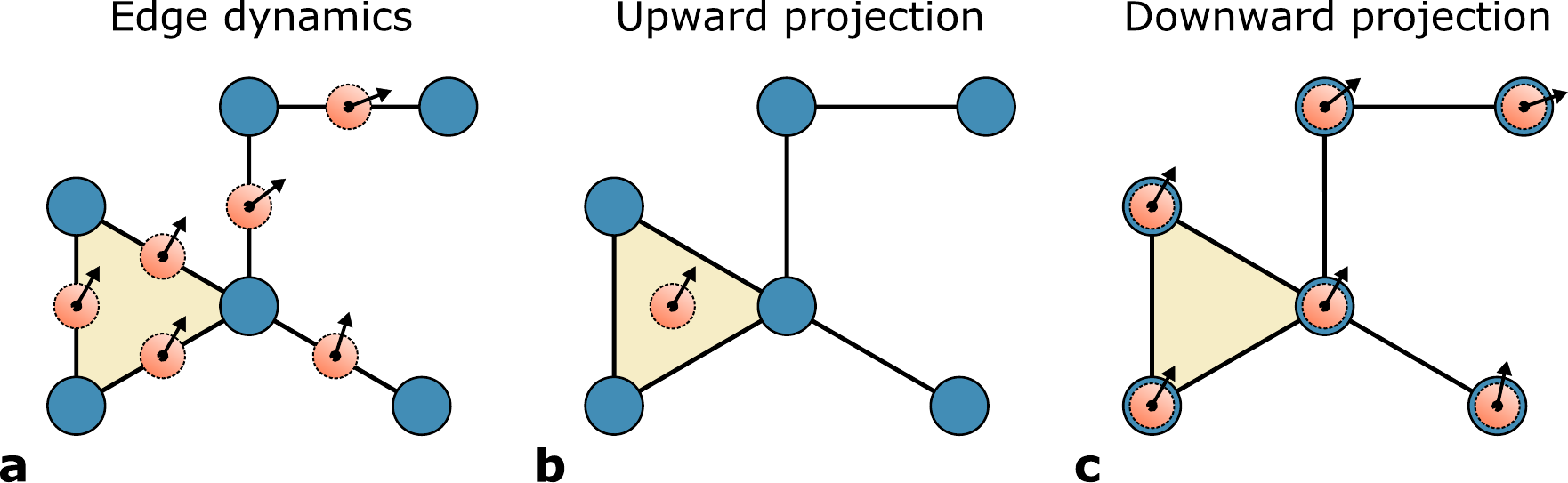}
	\caption{\textbf{Higher-order systems are fully dynamical.}
		{\bf a,} As opposed to traditional descriptions focused on node dynamics, it is possible to define state variables for hyperedges or simplices of arbitrary order, for example by associating oscillators to edges \cite{millan2020explosive} and coupling them to each other using their higher-order adjacency. In so doing, the distinction between dynamical units and interactions dissolves, and dependencies and feedback loops between orders become possible. {\bf b,} For example, it is possible to project the dynamics of hyperedges of order $k$ ($k=1$, edges or 1-simplices) onto their analogues of larger order ($k=2$, 3-hyperedges or 2-simplices). {\bf c,} It is similarly possible to project onto analogues of smaller order ($k=0$, nodes).}
	\label{fig:fig3}
\end{figure*}

Explosiveness is not limited to spreading processes. Of paramount importance for biology and neuroscience are systems of coupled oscillators, where the states of the nodes are $d$-dimensional continuous variables that evolve over time under mutual influence (Fig. \ref{fig:fig2}b). The most well-known setup is probably the one introduced by Yoshiki Kuramoto~\cite{kuramoto1975self}, in which unidimensional phase oscillators are endowed with natural frequencies, and interactions occur through sinusoidal couplings. When generalized to account for structured higher-order interactions among oscillators, the additional non-linearity generates abrupt switches between synchronized and incoherent states~\cite{skardal2020higher}. The emergence of bistability and the appearance of hysteresis cycles are driven by the presence of higher-order interactions alone, without the need for ad-hoc coupling mechanisms between the dynamical evolution and the local connectivity of the nodes.

In both examples, the introduction of higher-order interactions corresponds to having the state variable of a node influenced by a non-linear combination of the states of several other nodes. Tuning the relative importance of the strength of the higher-order and pairwise interactions provides a way in both cases
to change the nature of the transition from continuous to discontinuous
(Fig. \ref{fig:fig2}c). The similarity of the mechanisms yielding a first-order transition in these two very different dynamical processes leads to conjecture that the introduction of non-linear higher-order interactions and the tuning of their intensity is a general ingredient sufficient to provide abrupt transitions in a dynamical process. 

Despite this preliminary evidence, however, a rigorous and general proof of this conjecture is still lacking. Approximate approaches based on linearization around a fixed point of ordinary differential equations link the stability of hypergraphs dynamics to their graph projections~\cite{lucas2020multiorder, gambuzza2021stability, zhang2021generalized}, suggesting general conditions for stability associated to the different orders of the interactions. Mean-field treatment allows for an analytical solution for diffusion and spreading processes on arbitrary structures, separating stability conditions into structural and dynamical terms~\cite{de2021phase, stonge2020master}. A general argument based on bifurcation theory shows that variations on pairwise models, such as adding higher-order interactions, can lead to a change of critical behaviour from a continuous to a discontinuous transition for a wide class of models, including epidemic, synchronization and percolation transitions~\cite{kuehn2021universal}. Under some conditions, mathematicians have been able to formally prove that higher-order interactions are sufficient to induce bi-stable behaviour in the Susceptible-Infected-Susceptible (SIS) model, whereas it is impossible to achieve bistability in the traditional pairwise scheme~\cite{cisneros2020multi}. All in all, findings indicate that the presence of higher-order interactions provides a general pathway to explosive phenomena.
Yet, this marker of fragility of the collective behaviour in higher-order systems is still awaiting formal proof.\\

{\bf Topological dynamical processes.}
Most of the research on dynamical processes on networks has focused on the dynamics of node states, with interactions mediated by links. 
This is a natural and intuitive approach, because it describes the evolution of the most basic units of the system, coupled through the only possible (and simplest) interactions in networks~\cite{porter2016dynamical}. But by encoding higher-order interactions, it becomes possible to define couplings between interactions of different orders (nodes, and hyperedges or simplices). More importantly, we can associate state variables, not only to nodes, but also to hyperedges and simplices. For example, the state of an edge can influence the states of its two associated nodes, while contributing to and being influenced by the states of the higher-order interactions (for example, a 3-hyperedge) to which it belongs. In this way, higher-order dynamical systems transform static interactions into active agents that are coupled to the rest of the system and evolving in time. 

Recent results on simplicial oscillators offer a particularly striking example of this phenomenon. 
Consider a Kuramoto model defined on a simplicial complex comprising nodes, edges and 2-simplices (Fig. \ref{fig:fig3}a). 
In this case, phases are defined not only on nodes --- as in the traditional description --- but also on higher-order faces. 
The equations used in the classical formalism can be directly adapted to higher-order interactions, by substituting node incidence matrices with the appropriate higher-order analogues \cite{barahona2002synchronization}. 
In simplicial complexes, these matrices correspond to boundary operators between interactions of orders differing by one --- for example, node and edges, or edges and 2-simplices --- effectively providing a canonical mapping between phase dynamics of different orders.
Interesting phenomena emerge without adding further complications: 
the dynamics on 1-simplices (edges) displays a synchronization transition \cite{millan2020explosive} that is only revealed when projected onto simplices of higher (2-simplices, Fig. \ref{fig:fig3}b) or lower (nodes, Fig. \ref{fig:fig3}c) dimension. Indeed, phase transitions appear in both projected dynamics. 
And when the dynamics of the $(n-1)$- and $(n+1)$-simplices are coupled via the respective global order parameters, these transitions become explosive. 

The Hodge decomposition provides a rationale for this behaviour in terms of the inner structure of higher-order states \cite{schaub2020random}.  
In fact, these can be decomposed into harmonic, solenoidal and irrotational components, corresponding respectively to the dynamics induced by the kernel of the higher-order Laplacian, and to those induced by the projection to simplices one dimension higher and lower. 
In this light, higher-order systems can be considered collections of topological signals --- time series associated with interactions on all orders, which lend themselves to analysis using tools at the interface between algebraic topology, differential geometry and discrete calculus \cite{ghorbanchian2020higher,schaub2020random}.
As an example of this paradigm, higher-order Laplacians were recently shown to improve the description of flow information on edges with respect to standard graph Laplacians \cite{schaub2018flow}. The description was improved even when simplicial complexes contained only nodes and edges. Higher-order Laplacians also provided the first formulation for signal processing on generic topological spaces\cite{barbarossa2020topological}.  

Finally, even when states for higher-order interactions are defined, the topological structure of the system --- that is, the presence or absence of simplices and hyperedges --- has typically been considered fixed in time (for example, in neural codes \cite{ganmor2011sparse}).
However, in many systems the organization of the interactions changes over time \cite{holme2012temporal}. It remains an open question how to define realistic models of topological co-evolution, where higher-order structure and higher-order dynamics evolve together under the effect of mutual feedback \cite{gross2008adaptive}. \\

\begin{figure}
\centering
\includegraphics[width=0.5\textwidth]{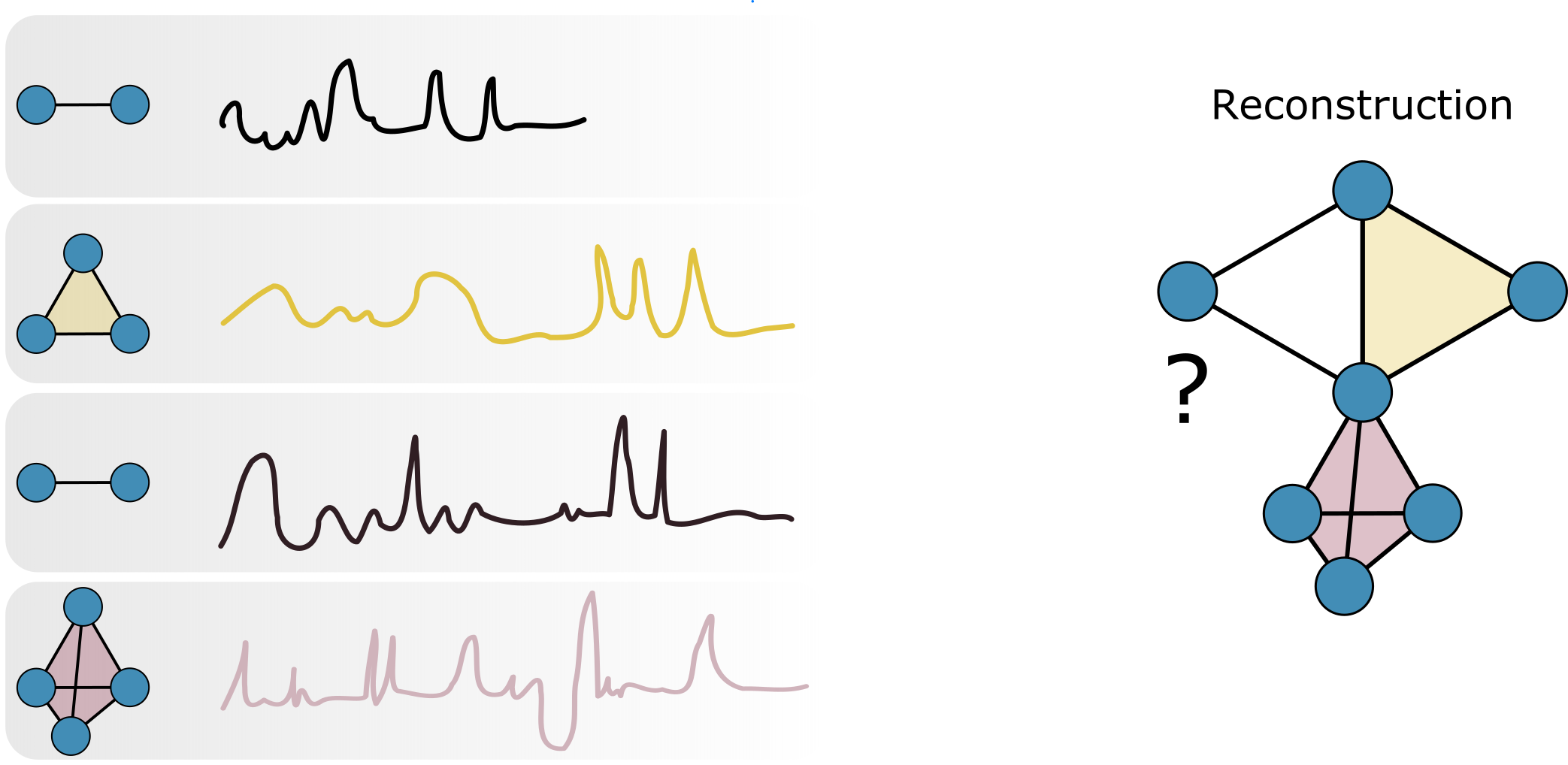}
\caption{\textbf{Inference of higher-order systems is still an open and challenging problem.} 
In spite of abundant network data, few records contain the information necessary to reconstruct a system's higher-order interactions. A number of tools and concepts have been proposed to overcome this problem, but existing methods to extract signals associated with higher-order interactions are still lacking. Reconstruction techniques based on a combination of data-driven modelling and Bayesian inference offer early evidence of an effective approach.   
}
\label{fig:fig4}
\end{figure}

{\bf Inferring higher-order interactions from data.}
A crucial ingredient in modelling real systems is the reconstruction of higher-order interactions from data (Fig. \ref{fig:fig4}).  
The vast majority of data available on network systems contains only records of pairwise interactions, even when the underlying rules rely on higher-order patterns. 
Naively attributing every observed dense subgraph in the
pairwise network (for example. triangles and larger cliques) to a putative higher-order interaction conflates the existence of an actual hyperedge with the coincidental accumulation of edges, which may otherwise emerge from community
structure, homophily or a geometric embedding. 
Recent work~\cite{young2020hypergraph} has demonstrated that it is possible to
distinguish between hyperedges and combinations of lower-order edges by
casting the problem as a Bayesian inference task, taking into account
the parsimony of the resulting reconstruction. 
With such an approach, hyperedges are identified only if they are supported by statistical evidence. 
It is as yet unclear how such approaches might be generalized to include more realistic modelling assumptions, containing a tighter interplay with mesoscale structures and latent space embeddings~\cite{saebi2020honem}.

Even when explicit hyperedge data are available, just as with pairwise network data, errors and incompleteness are unavoidable, requiring us to reconstruct the object of study from uncertain observations~\cite{newman2018structure,peixoto2018network}. 
For hypergraph data, recent work~\cite{musciotto2021detecting} has proposed an approach based on comparisons with null models, which is capable of
filtering out hyperedges that are not statistically significant. 
More work is needed to provide uncertainty quantification on the analyses that are conditioned on the reconstruction, as well as leveraging
more advanced techniques of hyperedge prediction to improve accuracy.

In addition to the reconstruction from direct but uncertain data, it is challenging to infer higher-order structures from indirect data such as time series, which encode the dynamical behaviour of the nodes rather than directly measured edges and hyperedges. 
This is an important issue in many biological systems such as the brain, where diseases like Parkinson's and schizophrenia have been associated with
dysfunctional brain connectivity~\cite{bohland2009proposal,bassett2017network,glomb2021computational}, but direct network measurements are often not available. 
A common approach is to compute correlations~\cite{skudlarski2008measuring} and measure synchronization~\cite{tass1998detection} between time
series. 
However, these approaches yield an unreliable understanding of the underlying system, because they cannot distinguish correlation from causation --- two or more nodes can be highly correlated even if they do not share an edge or hyperedge. Another set of approaches involves
exploiting temporal correlations, for example, the phase-dynamics reconstruction given a set of multivariate time series~\cite{kralemann2008phase}. Originally devised for pairwise interactions only, this methodology has been generalized to account for small motifs of interacting units~\cite{kralemann2011reconstructing}. 

The development of new synchronization measures for triplets has made it possible to identify multi-body locking from experimental data, even when every pair of systems taken in isolation remains asynchronous~\cite{kralemann2013detecting}. 
This approach can better differentiate between the physical connections and the effective ones, which are associated with the temporal influence of one node on another, leading to more reliable network reconstruction methods~\cite{kralemann2014reconstructing, pikovsky2018reconstruction}.

Finally, another possibility involves extending information-theoretical techniques, such as Granger causality~\cite{barnett2009granger} and transfer entropy~\cite{schreiber2000measuring}, to account for the existence of multi-body interactions.  Despite promising first steps in reconstructing higher-order interactions from static lower-order projections \cite{young2020hypergraph} and in multi-body information-theoretic quantities \cite{rosas2019quantifying}, the task of broadening this framework to consider fully higher-order interaction schemes remains an open problem. 

Reconstruction methods that are based only on temporal correlations still suffer from the problem of not being able to fully distinguish between direct and indirect causation, meaning they cannot differentiate between the existence of an actual edge or hyperedge between nodes and a longer path that connects them. They are similarly incapable of discerning non-causal correlation. 
Circumventing this problem is only possible in general if we can make interventions, rather than relying on observational data alone~\cite{pearl2009causality}. Nevertheless, methods based on Bayesian inference of generative models are able to convey the uncertainty about the causal relationships~\cite{peixoto2019network}. An important future direction is to generalize such methods to incorporate higher-order interactions~\cite{bianconi2014triadic,courtney2016generalized,chodrow2020configuration,chodrow2021generative,courtney2017weighted,kovalenko2021growing,peixoto2021disentangling} that vary in time~\cite{cencetti2021temporal} and describe emergent higher-order geometry~\cite{bianconi2017emergent}.\\

{\bf Future directions from past inspiration.}
The study of networked systems with higher-order interactions is still in its infancy, posing new challenges and opportunities for discoveries~\cite{battiston2020networks, torres2020representations, bick2021higher, benson2021higher}. Yet, it is also inspired by ideas from the past. For example, earlier work considered systems of coupled cells where dependencies of different orders were encoded via particular graph structures~\cite{golubitsky2006nonlinear}, clarifying how higher-order symmetries affect synchronization~\cite{stewart2003symmetry, golubitsky2005patterns}. 
Higher-order interactions can also generate new insights on older problems where they emerge as effective theories. A paradigmatic example is that of networks of phase oscillators with higher-order interactions, which arise from the phase reduction of nonlinear oscillator systems~\cite{nakao2016phase, pietras2019network, matheny2019exotic}. As a consequence, understanding the dynamics of phase-reduced systems with higher-order interactions can also clarify the physics of the general higher-dimensional system~\cite{komarov2013dynamics, ashwin2016hopf, ashwin2016identical,leon2019phase}, in particular at the onset of chaos~\cite{bick2016chaos} and metastable chimeras~\cite{bick2018heteroclinic, bick2019heteroclinic}.
Thus, in addition to providing an exciting way forward for network science, higher-order interactions can also create opportunities for a wider dialogue on the physics of dynamical systems.

From $p$-spin models~\cite{kirkpatrick1987dynamics,yu2011higher} to multilayer~\cite{de2016physics} and non-Markovian temporal networks~\cite{lambiotte2019networks}, the past suggests that new phenomena may occur when more realistic patterns of interactions are considered. 
Overcoming previous limitations, new data and new theory are now informing our network models beyond pairwise interactions~\cite{battiston2020networks}.


\begin{thebibliography}{111}%
	\makeatletter
	\providecommand \@ifxundefined [1]{%
		\@ifx{#1\undefined}
	}%
	\providecommand \@ifnum [1]{%
		\ifnum #1\expandafter \@firstoftwo
		\else \expandafter \@secondoftwo
		\fi
	}%
	\providecommand \@ifx [1]{%
		\ifx #1\expandafter \@firstoftwo
		\else \expandafter \@secondoftwo
		\fi
	}%
	\providecommand \natexlab [1]{#1}%
	\providecommand \enquote  [1]{``#1''}%
	\providecommand \bibnamefont  [1]{#1}%
	\providecommand \bibfnamefont [1]{#1}%
	\providecommand \citenamefont [1]{#1}%
	\providecommand \href@noop [0]{\@secondoftwo}%
	\providecommand \href [0]{\begingroup \@sanitize@url \@href}%
	\providecommand \@href[1]{\@@startlink{#1}\@@href}%
	\providecommand \@@href[1]{\endgroup#1\@@endlink}%
	\providecommand \@sanitize@url [0]{\catcode `\\12\catcode `\$12\catcode
		`\&12\catcode `\#12\catcode `\^12\catcode `\_12\catcode `\%12\relax}%
	\providecommand \@@startlink[1]{}%
	\providecommand \@@endlink[0]{}%
	\providecommand \url  [0]{\begingroup\@sanitize@url \@url }%
	\providecommand \@url [1]{\endgroup\@href {#1}{\urlprefix }}%
	\providecommand \urlprefix  [0]{URL }%
	\providecommand \Eprint [0]{\href }%
	\providecommand \doibase [0]{http://dx.doi.org/}%
	\providecommand \selectlanguage [0]{\@gobble}%
	\providecommand \bibinfo  [0]{\@secondoftwo}%
	\providecommand \bibfield  [0]{\@secondoftwo}%
	\providecommand \translation [1]{[#1]}%
	\providecommand \BibitemOpen [0]{}%
	\providecommand \bibitemStop [0]{}%
	\providecommand \bibitemNoStop [0]{.\EOS\space}%
	\providecommand \EOS [0]{\spacefactor3000\relax}%
	\providecommand \BibitemShut  [1]{\csname bibitem#1\endcsname}%
	\let\auto@bib@innerbib\@empty
	\bibitem [{\citenamefont {Barab{\'a}si}(2011)}]{barabasi2011network}%
	\BibitemOpen
	\bibfield  {author} {\bibinfo {author} {\bibfnamefont {A.-L.}\ \bibnamefont
			{Barab{\'a}si}},\ }\href@noop {} {\bibfield  {journal} {\bibinfo  {journal}
			{Nat. Phys.}\ }\textbf {\bibinfo {volume} {8}},\ \bibinfo {pages} {14}
		(\bibinfo {year} {2011})}\BibitemShut {NoStop}%
	\bibitem [{\citenamefont {Boccaletti}\ \emph {et~al.}(2006)\citenamefont
		{Boccaletti}, \citenamefont {Latora}, \citenamefont {Moreno}, \citenamefont
		{Chavez},\ and\ \citenamefont {Hwang}}]{boccaletti2006complex}%
	\BibitemOpen
	\bibfield  {author} {\bibinfo {author} {\bibfnamefont {S.}~\bibnamefont
			{Boccaletti}}, \bibinfo {author} {\bibfnamefont {V.}~\bibnamefont {Latora}},
		\bibinfo {author} {\bibfnamefont {Y.}~\bibnamefont {Moreno}}, \bibinfo
		{author} {\bibfnamefont {M.}~\bibnamefont {Chavez}}, \ and\ \bibinfo {author}
		{\bibfnamefont {D.-U.}\ \bibnamefont {Hwang}},\ }\href@noop {} {\bibfield
		{journal} {\bibinfo  {journal} {Phys. Rep.}\ }\textbf {\bibinfo {volume}
			{424}},\ \bibinfo {pages} {175} (\bibinfo {year} {Fervier 2006})}\BibitemShut
	{NoStop}%
	\bibitem [{\citenamefont {Barrat}\ \emph {et~al.}(2008)\citenamefont {Barrat},
		\citenamefont {Barthelemy},\ and\ \citenamefont
		{Vespignani}}]{barrat2008dynamical}%
	\BibitemOpen
	\bibfield  {author} {\bibinfo {author} {\bibfnamefont {A.}~\bibnamefont
			{Barrat}}, \bibinfo {author} {\bibfnamefont {M.}~\bibnamefont {Barthelemy}},
		\ and\ \bibinfo {author} {\bibfnamefont {A.}~\bibnamefont {Vespignani}},\
	}\href@noop {} {\emph {\bibinfo {title} {Dynamical processes on complex
				networks}}}\ (\bibinfo  {publisher} {Cambridge university press},\ \bibinfo
	{year} {2008})\BibitemShut {NoStop}%
	\bibitem [{\citenamefont {Vespignani}(2012)}]{vespignani2012modelling}%
	\BibitemOpen
	\bibfield  {author} {\bibinfo {author} {\bibfnamefont {A.}~\bibnamefont
			{Vespignani}},\ }\href@noop {} {\bibfield  {journal} {\bibinfo  {journal}
			{Nat. Phys.}\ }\textbf {\bibinfo {volume} {8}},\ \bibinfo {pages} {32}
		(\bibinfo {year} {2012})}\BibitemShut {NoStop}%
	\bibitem [{\citenamefont {{Pastor-Satorras}}\ and\ \citenamefont
		{Vespignani}(2001)}]{pastor2001epidemic}%
	\BibitemOpen
	\bibfield  {author} {\bibinfo {author} {\bibfnamefont {R.}~\bibnamefont
			{{Pastor-Satorras}}}\ and\ \bibinfo {author} {\bibfnamefont {A.}~\bibnamefont
			{Vespignani}},\ }\href@noop {} {\bibfield  {journal} {\bibinfo  {journal}
			{Phys. Rev. Lett.}\ }\textbf {\bibinfo {volume} {86}},\ \bibinfo {pages}
		{3200} (\bibinfo {year} {2001})}\BibitemShut {NoStop}%
	\bibitem [{\citenamefont {Bogun{\'a}}\ \emph {et~al.}(2003)\citenamefont
		{Bogun{\'a}}, \citenamefont {Pastor-Satorras},\ and\ \citenamefont
		{Vespignani}}]{boguna2003absence}%
	\BibitemOpen
	\bibfield  {author} {\bibinfo {author} {\bibfnamefont {M.}~\bibnamefont
			{Bogun{\'a}}}, \bibinfo {author} {\bibfnamefont {R.}~\bibnamefont
			{Pastor-Satorras}}, \ and\ \bibinfo {author} {\bibfnamefont {A.}~\bibnamefont
			{Vespignani}},\ }\href@noop {} {\bibfield  {journal} {\bibinfo  {journal}
			{Phys. Rev. Lett.}\ }\textbf {\bibinfo {volume} {90}},\ \bibinfo {pages}
		{028701} (\bibinfo {year} {2003})}\BibitemShut {NoStop}%
	\bibitem [{\citenamefont {Barahona}\ and\ \citenamefont
		{Pecora}(2002)}]{barahona2002synchronization}%
	\BibitemOpen
	\bibfield  {author} {\bibinfo {author} {\bibfnamefont {M.}~\bibnamefont
			{Barahona}}\ and\ \bibinfo {author} {\bibfnamefont {L.~M.}\ \bibnamefont
			{Pecora}},\ }\href@noop {} {\bibfield  {journal} {\bibinfo  {journal} {Phys.
				Rev. Lett.}\ }\textbf {\bibinfo {volume} {89}},\ \bibinfo {pages} {054101}
		(\bibinfo {year} {2002})}\BibitemShut {NoStop}%
	\bibitem [{\citenamefont {Del~Vicario}\ \emph {et~al.}(2016)\citenamefont
		{Del~Vicario}, \citenamefont {Bessi}, \citenamefont {Zollo}, \citenamefont
		{Petroni}, \citenamefont {Scala}, \citenamefont {Caldarelli}, \citenamefont
		{Stanley},\ and\ \citenamefont {Quattrociocchi}}]{del2016spreading}%
	\BibitemOpen
	\bibfield  {author} {\bibinfo {author} {\bibfnamefont {M.}~\bibnamefont
			{Del~Vicario}}, \bibinfo {author} {\bibfnamefont {A.}~\bibnamefont {Bessi}},
		\bibinfo {author} {\bibfnamefont {F.}~\bibnamefont {Zollo}}, \bibinfo
		{author} {\bibfnamefont {F.}~\bibnamefont {Petroni}}, \bibinfo {author}
		{\bibfnamefont {A.}~\bibnamefont {Scala}}, \bibinfo {author} {\bibfnamefont
			{G.}~\bibnamefont {Caldarelli}}, \bibinfo {author} {\bibfnamefont {H.~E.}\
			\bibnamefont {Stanley}}, \ and\ \bibinfo {author} {\bibfnamefont
			{W.}~\bibnamefont {Quattrociocchi}},\ }\href@noop {} {\bibfield  {journal}
		{\bibinfo  {journal} {Proceedings of the National Academy of Sciences}\
		}\textbf {\bibinfo {volume} {113}},\ \bibinfo {pages} {554} (\bibinfo {year}
		{2016})}\BibitemShut {NoStop}%
	\bibitem [{\citenamefont {Patania}\ \emph {et~al.}(2017)\citenamefont
		{Patania}, \citenamefont {Vaccarino},\ and\ \citenamefont
		{Petri}}]{patania2017topological}%
	\BibitemOpen
	\bibfield  {author} {\bibinfo {author} {\bibfnamefont {A.}~\bibnamefont
			{Patania}}, \bibinfo {author} {\bibfnamefont {F.}~\bibnamefont {Vaccarino}},
		\ and\ \bibinfo {author} {\bibfnamefont {G.}~\bibnamefont {Petri}},\
	}\href@noop {} {\bibfield  {journal} {\bibinfo  {journal} {EPJ Data Sci.}\
		}\textbf {\bibinfo {volume} {6}},\ \bibinfo {pages} {7} (\bibinfo {year}
		{2017})}\BibitemShut {NoStop}%
	\bibitem [{\citenamefont {Battiston}\ \emph {et~al.}(2020)\citenamefont
		{Battiston}, \citenamefont {Cencetti}, \citenamefont {Iacopini},
		\citenamefont {Latora}, \citenamefont {Lucas}, \citenamefont {Patania},
		\citenamefont {Young},\ and\ \citenamefont {Petri}}]{battiston2020networks}%
	\BibitemOpen
	\bibfield  {author} {\bibinfo {author} {\bibfnamefont {F.}~\bibnamefont
			{Battiston}}, \bibinfo {author} {\bibfnamefont {G.}~\bibnamefont {Cencetti}},
		\bibinfo {author} {\bibfnamefont {I.}~\bibnamefont {Iacopini}}, \bibinfo
		{author} {\bibfnamefont {V.}~\bibnamefont {Latora}}, \bibinfo {author}
		{\bibfnamefont {M.}~\bibnamefont {Lucas}}, \bibinfo {author} {\bibfnamefont
			{A.}~\bibnamefont {Patania}}, \bibinfo {author} {\bibfnamefont {J.-G.}\
			\bibnamefont {Young}}, \ and\ \bibinfo {author} {\bibfnamefont
			{G.}~\bibnamefont {Petri}},\ }\href@noop {} {\bibfield  {journal} {\bibinfo
			{journal} {Phys. Rep.}\ }\textbf {\bibinfo {volume} {874}},\ \bibinfo {pages}
		{1} (\bibinfo {year} {2020})}\BibitemShut {NoStop}%
	\bibitem [{\citenamefont {Schneidman}\ \emph {et~al.}(2006)\citenamefont
		{Schneidman}, \citenamefont {Berry~II}, \citenamefont {Segev},\ and\
		\citenamefont {Bialek}}]{schneidman2006weak}%
	\BibitemOpen
	\bibfield  {author} {\bibinfo {author} {\bibfnamefont {E.}~\bibnamefont
			{Schneidman}}, \bibinfo {author} {\bibfnamefont {M.~J.}\ \bibnamefont
			{Berry~II}}, \bibinfo {author} {\bibfnamefont {R.}~\bibnamefont {Segev}}, \
		and\ \bibinfo {author} {\bibfnamefont {W.}~\bibnamefont {Bialek}},\
	}\href@noop {} {\bibfield  {journal} {\bibinfo  {journal} {Nature}\ }\textbf
		{\bibinfo {volume} {440}},\ \bibinfo {pages} {1007} (\bibinfo {year}
		{2006})}\BibitemShut {NoStop}%
	\bibitem [{\citenamefont {Schneidman}\ \emph {et~al.}(2003)\citenamefont
		{Schneidman}, \citenamefont {Still}, \citenamefont {Berry}, \citenamefont
		{Bialek} \emph {et~al.}}]{schneidman2003network}%
	\BibitemOpen
	\bibfield  {author} {\bibinfo {author} {\bibfnamefont {E.}~\bibnamefont
			{Schneidman}}, \bibinfo {author} {\bibfnamefont {S.}~\bibnamefont {Still}},
		\bibinfo {author} {\bibfnamefont {M.~J.}\ \bibnamefont {Berry}}, \bibinfo
		{author} {\bibfnamefont {W.}~\bibnamefont {Bialek}},  \emph {et~al.},\
	}\href@noop {} {\bibfield  {journal} {\bibinfo  {journal} {Phys. Rev. Lett.}\
		}\textbf {\bibinfo {volume} {91}},\ \bibinfo {pages} {238701} (\bibinfo
		{year} {2003})}\BibitemShut {NoStop}%
	\bibitem [{\citenamefont {Yu}\ \emph {et~al.}(2011)\citenamefont {Yu},
		\citenamefont {Yang}, \citenamefont {Nakahara}, \citenamefont {Santos},
		\citenamefont {Nikoli{\'c}},\ and\ \citenamefont {Plenz}}]{yu2011higher}%
	\BibitemOpen
	\bibfield  {author} {\bibinfo {author} {\bibfnamefont {S.}~\bibnamefont
			{Yu}}, \bibinfo {author} {\bibfnamefont {H.}~\bibnamefont {Yang}}, \bibinfo
		{author} {\bibfnamefont {H.}~\bibnamefont {Nakahara}}, \bibinfo {author}
		{\bibfnamefont {G.~S.}\ \bibnamefont {Santos}}, \bibinfo {author}
		{\bibfnamefont {D.}~\bibnamefont {Nikoli{\'c}}}, \ and\ \bibinfo {author}
		{\bibfnamefont {D.}~\bibnamefont {Plenz}},\ }\href@noop {} {\bibfield
		{journal} {\bibinfo  {journal} {J. Neurosci.}\ }\textbf {\bibinfo {volume}
			{31}},\ \bibinfo {pages} {17514} (\bibinfo {year} {2011})}\BibitemShut
	{NoStop}%
	\bibitem [{\citenamefont {Giusti}\ \emph {et~al.}(2015)\citenamefont {Giusti},
		\citenamefont {Pastalkova}, \citenamefont {Curto},\ and\ \citenamefont
		{Itskov}}]{giusti2015clique}%
	\BibitemOpen
	\bibfield  {author} {\bibinfo {author} {\bibfnamefont {C.}~\bibnamefont
			{Giusti}}, \bibinfo {author} {\bibfnamefont {E.}~\bibnamefont {Pastalkova}},
		\bibinfo {author} {\bibfnamefont {C.}~\bibnamefont {Curto}}, \ and\ \bibinfo
		{author} {\bibfnamefont {V.}~\bibnamefont {Itskov}},\ }\href@noop {}
	{\bibfield  {journal} {\bibinfo  {journal} {Proc. Natl. Acad. Sci. U.S.A.}\
		}\textbf {\bibinfo {volume} {112}},\ \bibinfo {pages} {13455} (\bibinfo
		{year} {2015})}\BibitemShut {NoStop}%
	\bibitem [{\citenamefont {Gardner}\ \emph {et~al.}(2021)\citenamefont
		{Gardner}, \citenamefont {Hermansen}, \citenamefont {Pachitariu},
		\citenamefont {Burak}, \citenamefont {Baas}, \citenamefont {Dunn},
		\citenamefont {Moser},\ and\ \citenamefont {Moser}}]{gardner2021toroidal}%
	\BibitemOpen
	\bibfield  {author} {\bibinfo {author} {\bibfnamefont {R.~J.}\ \bibnamefont
			{Gardner}}, \bibinfo {author} {\bibfnamefont {E.}~\bibnamefont {Hermansen}},
		\bibinfo {author} {\bibfnamefont {M.}~\bibnamefont {Pachitariu}}, \bibinfo
		{author} {\bibfnamefont {Y.}~\bibnamefont {Burak}}, \bibinfo {author}
		{\bibfnamefont {N.~A.}\ \bibnamefont {Baas}}, \bibinfo {author}
		{\bibfnamefont {B.~J.}\ \bibnamefont {Dunn}}, \bibinfo {author}
		{\bibfnamefont {M.-B.}\ \bibnamefont {Moser}}, \ and\ \bibinfo {author}
		{\bibfnamefont {E.~I.}\ \bibnamefont {Moser}},\ }\href@noop {} {\bibfield
		{journal} {\bibinfo  {journal} {bioRxiv}\ } (\bibinfo {year}
		{2021})}\BibitemShut {NoStop}%
	\bibitem [{\citenamefont {Ganmor}\ \emph {et~al.}(2011)\citenamefont {Ganmor},
		\citenamefont {Segev},\ and\ \citenamefont {Schneidman}}]{ganmor2011sparse}%
	\BibitemOpen
	\bibfield  {author} {\bibinfo {author} {\bibfnamefont {E.}~\bibnamefont
			{Ganmor}}, \bibinfo {author} {\bibfnamefont {R.}~\bibnamefont {Segev}}, \
		and\ \bibinfo {author} {\bibfnamefont {E.}~\bibnamefont {Schneidman}},\
	}\href@noop {} {\bibfield  {journal} {\bibinfo  {journal} {Proc. Natl. Acad.
				Sci. U.S.A.}\ }\textbf {\bibinfo {volume} {108}},\ \bibinfo {pages} {9679}
		(\bibinfo {year} {2011})}\BibitemShut {NoStop}%
	\bibitem [{\citenamefont {Merchan}\ and\ \citenamefont
		{Nemenman}(2016)}]{merchan2016sufficiency}%
	\BibitemOpen
	\bibfield  {author} {\bibinfo {author} {\bibfnamefont {L.}~\bibnamefont
			{Merchan}}\ and\ \bibinfo {author} {\bibfnamefont {I.}~\bibnamefont
			{Nemenman}},\ }\href@noop {} {\bibfield  {journal} {\bibinfo  {journal}
			{Journal of Statistical Physics}\ }\textbf {\bibinfo {volume} {162}},\
		\bibinfo {pages} {1294} (\bibinfo {year} {2016})}\BibitemShut {NoStop}%
	\bibitem [{\citenamefont {Mayfield}\ and\ \citenamefont
		{Stouffer}(2017)}]{mayfield2017higher}%
	\BibitemOpen
	\bibfield  {author} {\bibinfo {author} {\bibfnamefont {M.~M.}\ \bibnamefont
			{Mayfield}}\ and\ \bibinfo {author} {\bibfnamefont {D.~B.}\ \bibnamefont
			{Stouffer}},\ }\href@noop {} {\bibfield  {journal} {\bibinfo  {journal} {Nat.
				Ecol. Evol.}\ }\textbf {\bibinfo {volume} {1}},\ \bibinfo {pages} {0062}
		(\bibinfo {year} {2017})}\BibitemShut {NoStop}%
	\bibitem [{\citenamefont {Cervantes-Loreto}\ \emph {et~al.}(2020)\citenamefont
		{Cervantes-Loreto}, \citenamefont {Ayers}, \citenamefont {Dobbs},
		\citenamefont {Brosi},\ and\ \citenamefont
		{Stouffer}}]{cervantes2020context}%
	\BibitemOpen
	\bibfield  {author} {\bibinfo {author} {\bibfnamefont {A.}~\bibnamefont
			{Cervantes-Loreto}}, \bibinfo {author} {\bibfnamefont {C.}~\bibnamefont
			{Ayers}}, \bibinfo {author} {\bibfnamefont {E.}~\bibnamefont {Dobbs}},
		\bibinfo {author} {\bibfnamefont {B.}~\bibnamefont {Brosi}}, \ and\ \bibinfo
		{author} {\bibfnamefont {D.}~\bibnamefont {Stouffer}},\ }\href@noop {}
	{\bibfield  {journal} {\bibinfo  {journal} {Authorea Preprints}\ } (\bibinfo
		{year} {2020})}\BibitemShut {NoStop}%
	\bibitem [{\citenamefont {Bairey}\ \emph {et~al.}(2017)\citenamefont {Bairey},
		\citenamefont {Kelsic},\ and\ \citenamefont {Kishony}}]{bairey2016high}%
	\BibitemOpen
	\bibfield  {author} {\bibinfo {author} {\bibfnamefont {E.}~\bibnamefont
			{Bairey}}, \bibinfo {author} {\bibfnamefont {E.~D.}\ \bibnamefont {Kelsic}},
		\ and\ \bibinfo {author} {\bibfnamefont {R.}~\bibnamefont {Kishony}},\
	}\href@noop {} {\bibfield  {journal} {\bibinfo  {journal} {Nat. Commun.}\
		}\textbf {\bibinfo {volume} {7}},\ \bibinfo {pages} {12285} (\bibinfo {year}
		{2017})}\BibitemShut {NoStop}%
	\bibitem [{\citenamefont {Grilli}\ \emph {et~al.}(2017)\citenamefont {Grilli},
		\citenamefont {Barab{\'a}s}, \citenamefont {{Michalska-Smith}},\ and\
		\citenamefont {Allesina}}]{grilli2017higher}%
	\BibitemOpen
	\bibfield  {author} {\bibinfo {author} {\bibfnamefont {J.}~\bibnamefont
			{Grilli}}, \bibinfo {author} {\bibfnamefont {G.}~\bibnamefont {Barab{\'a}s}},
		\bibinfo {author} {\bibfnamefont {M.~J.}\ \bibnamefont {{Michalska-Smith}}},
		\ and\ \bibinfo {author} {\bibfnamefont {S.}~\bibnamefont {Allesina}},\
	}\href@noop {} {\bibfield  {journal} {\bibinfo  {journal} {Nature}\ }\textbf
		{\bibinfo {volume} {548}},\ \bibinfo {pages} {210} (\bibinfo {year}
		{2017})}\BibitemShut {NoStop}%
	\bibitem [{\citenamefont {Ritz}\ \emph {et~al.}(2014)\citenamefont {Ritz},
		\citenamefont {Tegge}, \citenamefont {Kim}, \citenamefont {Poirel},\ and\
		\citenamefont {Murali}}]{ritz2014signaling}%
	\BibitemOpen
	\bibfield  {author} {\bibinfo {author} {\bibfnamefont {A.}~\bibnamefont
			{Ritz}}, \bibinfo {author} {\bibfnamefont {A.~N.}\ \bibnamefont {Tegge}},
		\bibinfo {author} {\bibfnamefont {H.}~\bibnamefont {Kim}}, \bibinfo {author}
		{\bibfnamefont {C.~L.}\ \bibnamefont {Poirel}}, \ and\ \bibinfo {author}
		{\bibfnamefont {T.}~\bibnamefont {Murali}},\ }\href@noop {} {\bibfield
		{journal} {\bibinfo  {journal} {Trends Biotechnol.}\ }\textbf {\bibinfo
			{volume} {32}},\ \bibinfo {pages} {356} (\bibinfo {year} {2014})}\BibitemShut
	{NoStop}%
	\bibitem [{\citenamefont {Centola}\ \emph {et~al.}(2018)\citenamefont
		{Centola}, \citenamefont {Becker}, \citenamefont {Brackbill},\ and\
		\citenamefont {Baronchelli}}]{centola2018experimental}%
	\BibitemOpen
	\bibfield  {author} {\bibinfo {author} {\bibfnamefont {D.}~\bibnamefont
			{Centola}}, \bibinfo {author} {\bibfnamefont {J.}~\bibnamefont {Becker}},
		\bibinfo {author} {\bibfnamefont {D.}~\bibnamefont {Brackbill}}, \ and\
		\bibinfo {author} {\bibfnamefont {A.}~\bibnamefont {Baronchelli}},\
	}\href@noop {} {\bibfield  {journal} {\bibinfo  {journal} {Science}\ }\textbf
		{\bibinfo {volume} {360}},\ \bibinfo {pages} {1116} (\bibinfo {year}
		{2018})}\BibitemShut {NoStop}%
	\bibitem [{\citenamefont {Milojevi{\'c}}(2014)}]{milojevic2014principles}%
	\BibitemOpen
	\bibfield  {author} {\bibinfo {author} {\bibfnamefont {S.}~\bibnamefont
			{Milojevi{\'c}}},\ }\href@noop {} {\bibfield  {journal} {\bibinfo  {journal}
			{Proceedings of the National Academy of Sciences}\ }\textbf {\bibinfo
			{volume} {111}},\ \bibinfo {pages} {3984} (\bibinfo {year}
		{2014})}\BibitemShut {NoStop}%
	\bibitem [{\citenamefont {Povh}\ \emph {et~al.}(1995)\citenamefont {Povh},
		\citenamefont {Rith}, \citenamefont {Scholz},\ and\ \citenamefont
		{Zetsche}}]{povh1995many}%
	\BibitemOpen
	\bibfield  {author} {\bibinfo {author} {\bibfnamefont {B.}~\bibnamefont
			{Povh}}, \bibinfo {author} {\bibfnamefont {K.}~\bibnamefont {Rith}}, \bibinfo
		{author} {\bibfnamefont {C.}~\bibnamefont {Scholz}}, \ and\ \bibinfo {author}
		{\bibfnamefont {F.}~\bibnamefont {Zetsche}},\ }in\ \href@noop {} {\emph
		{\bibinfo {booktitle} {Particles and Nuclei}}}\ (\bibinfo  {publisher}
	{Springer},\ \bibinfo {year} {1995})\ pp.\ \bibinfo {pages}
	{281--303}\BibitemShut {NoStop}%
	\bibitem [{\citenamefont {Duck}(1966)}]{duck1966three}%
	\BibitemOpen
	\bibfield  {author} {\bibinfo {author} {\bibfnamefont {I.}~\bibnamefont
			{Duck}},\ }\href@noop {} {\bibfield  {journal} {\bibinfo  {journal} {Nuclear
				Physics}\ }\textbf {\bibinfo {volume} {84}},\ \bibinfo {pages} {586}
		(\bibinfo {year} {1966})}\BibitemShut {NoStop}%
	\bibitem [{\citenamefont {Kim}\ \emph {et~al.}(2006)\citenamefont {Kim},
		\citenamefont {Sofo}, \citenamefont {Velegol}, \citenamefont {Cole},\ and\
		\citenamefont {Lucas}}]{kim2006van}%
	\BibitemOpen
	\bibfield  {author} {\bibinfo {author} {\bibfnamefont {H.-Y.}\ \bibnamefont
			{Kim}}, \bibinfo {author} {\bibfnamefont {J.~O.}\ \bibnamefont {Sofo}},
		\bibinfo {author} {\bibfnamefont {D.}~\bibnamefont {Velegol}}, \bibinfo
		{author} {\bibfnamefont {M.~W.}\ \bibnamefont {Cole}}, \ and\ \bibinfo
		{author} {\bibfnamefont {A.~A.}\ \bibnamefont {Lucas}},\ }\href@noop {}
	{\bibfield  {journal} {\bibinfo  {journal} {J. Chem. Phys.}\ }\textbf
		{\bibinfo {volume} {124}},\ \bibinfo {pages} {074504} (\bibinfo {year}
		{2006})}\BibitemShut {NoStop}%
	\bibitem [{\citenamefont {Zeiher}\ \emph {et~al.}(2017)\citenamefont {Zeiher},
		\citenamefont {Choi}, \citenamefont {Rubio-Abadal}, \citenamefont {Pohl},
		\citenamefont {Van~Bijnen}, \citenamefont {Bloch},\ and\ \citenamefont
		{Gross}}]{zeiher2017coherent}%
	\BibitemOpen
	\bibfield  {author} {\bibinfo {author} {\bibfnamefont {J.}~\bibnamefont
			{Zeiher}}, \bibinfo {author} {\bibfnamefont {J.-y.}\ \bibnamefont {Choi}},
		\bibinfo {author} {\bibfnamefont {A.}~\bibnamefont {Rubio-Abadal}}, \bibinfo
		{author} {\bibfnamefont {T.}~\bibnamefont {Pohl}}, \bibinfo {author}
		{\bibfnamefont {R.}~\bibnamefont {Van~Bijnen}}, \bibinfo {author}
		{\bibfnamefont {I.}~\bibnamefont {Bloch}}, \ and\ \bibinfo {author}
		{\bibfnamefont {C.}~\bibnamefont {Gross}},\ }\href@noop {} {\bibfield
		{journal} {\bibinfo  {journal} {Phys. Rev. X}\ }\textbf {\bibinfo {volume}
			{7}},\ \bibinfo {pages} {041063} (\bibinfo {year} {2017})}\BibitemShut
	{NoStop}%
	\bibitem [{\citenamefont {Schaub}\ \emph {et~al.}(2020)\citenamefont {Schaub},
		\citenamefont {Benson}, \citenamefont {Horn}, \citenamefont {Lippner},\ and\
		\citenamefont {Jadbabaie}}]{schaub2020random}%
	\BibitemOpen
	\bibfield  {author} {\bibinfo {author} {\bibfnamefont {M.~T.}\ \bibnamefont
			{Schaub}}, \bibinfo {author} {\bibfnamefont {A.~R.}\ \bibnamefont {Benson}},
		\bibinfo {author} {\bibfnamefont {P.}~\bibnamefont {Horn}}, \bibinfo {author}
		{\bibfnamefont {G.}~\bibnamefont {Lippner}}, \ and\ \bibinfo {author}
		{\bibfnamefont {A.}~\bibnamefont {Jadbabaie}},\ }\href@noop {} {\bibfield
		{journal} {\bibinfo  {journal} {SIAM Rev.}\ }\textbf {\bibinfo {volume}
			{62}},\ \bibinfo {pages} {353} (\bibinfo {year} {2020})}\BibitemShut
	{NoStop}%
	\bibitem [{\citenamefont {Carletti}\ \emph {et~al.}(2020)\citenamefont
		{Carletti}, \citenamefont {Battiston}, \citenamefont {Cencetti},\ and\
		\citenamefont {Fanelli}}]{carletti2020random}%
	\BibitemOpen
	\bibfield  {author} {\bibinfo {author} {\bibfnamefont {T.}~\bibnamefont
			{Carletti}}, \bibinfo {author} {\bibfnamefont {F.}~\bibnamefont {Battiston}},
		\bibinfo {author} {\bibfnamefont {G.}~\bibnamefont {Cencetti}}, \ and\
		\bibinfo {author} {\bibfnamefont {D.}~\bibnamefont {Fanelli}},\ }\href@noop
	{} {\bibfield  {journal} {\bibinfo  {journal} {Phys. Rev. E}\ }\textbf
		{\bibinfo {volume} {101}},\ \bibinfo {pages} {022308} (\bibinfo {year}
		{2020})}\BibitemShut {NoStop}%
	\bibitem [{\citenamefont {Mill{\'a}n}\ \emph {et~al.}(2019)\citenamefont
		{Mill{\'a}n}, \citenamefont {Torres},\ and\ \citenamefont
		{Bianconi}}]{millan2019synchronization}%
	\BibitemOpen
	\bibfield  {author} {\bibinfo {author} {\bibfnamefont {A.~P.}\ \bibnamefont
			{Mill{\'a}n}}, \bibinfo {author} {\bibfnamefont {J.~J.}\ \bibnamefont
			{Torres}}, \ and\ \bibinfo {author} {\bibfnamefont {G.}~\bibnamefont
			{Bianconi}},\ }\href@noop {} {\bibfield  {journal} {\bibinfo  {journal}
			{Phys. Rev. E}\ }\textbf {\bibinfo {volume} {99}},\ \bibinfo {pages} {022307}
		(\bibinfo {year} {2019})}\BibitemShut {NoStop}%
	\bibitem [{\citenamefont {Skardal}\ and\ \citenamefont
		{Arenas}(2019)}]{skardal2019abrupt}%
	\BibitemOpen
	\bibfield  {author} {\bibinfo {author} {\bibfnamefont {P.~S.}\ \bibnamefont
			{Skardal}}\ and\ \bibinfo {author} {\bibfnamefont {A.}~\bibnamefont
			{Arenas}},\ }\href@noop {} {\bibfield  {journal} {\bibinfo  {journal} {Phys.
				Rev. Lett.}\ }\textbf {\bibinfo {volume} {122}},\ \bibinfo {pages} {248301}
		(\bibinfo {year} {2019})}\BibitemShut {NoStop}%
	\bibitem [{\citenamefont {Iacopini}\ \emph {et~al.}(2019)\citenamefont
		{Iacopini}, \citenamefont {Petri}, \citenamefont {Barrat},\ and\
		\citenamefont {Latora}}]{iacopini2019simplicial}%
	\BibitemOpen
	\bibfield  {author} {\bibinfo {author} {\bibfnamefont {I.}~\bibnamefont
			{Iacopini}}, \bibinfo {author} {\bibfnamefont {G.}~\bibnamefont {Petri}},
		\bibinfo {author} {\bibfnamefont {A.}~\bibnamefont {Barrat}}, \ and\ \bibinfo
		{author} {\bibfnamefont {V.}~\bibnamefont {Latora}},\ }\href@noop {}
	{\bibfield  {journal} {\bibinfo  {journal} {Nat. Commun.}\ }\textbf {\bibinfo
			{volume} {10}},\ \bibinfo {pages} {2485} (\bibinfo {year}
		{2019})}\BibitemShut {NoStop}%
	\bibitem [{\citenamefont {{de Arruda}}\ \emph {et~al.}(2020)\citenamefont {{de
				Arruda}}, \citenamefont {Petri},\ and\ \citenamefont
		{Moreno}}]{de2020social}%
	\BibitemOpen
	\bibfield  {author} {\bibinfo {author} {\bibfnamefont {G.~F.}\ \bibnamefont
			{{de Arruda}}}, \bibinfo {author} {\bibfnamefont {G.}~\bibnamefont {Petri}},
		\ and\ \bibinfo {author} {\bibfnamefont {Y.}~\bibnamefont {Moreno}},\
	}\href@noop {} {\bibfield  {journal} {\bibinfo  {journal} {Phys. Rev. Res.}\
		}\textbf {\bibinfo {volume} {2}},\ \bibinfo {pages} {023032} (\bibinfo {year}
		{2020})}\BibitemShut {NoStop}%
	\bibitem [{\citenamefont {Neuh{\"a}user}\ \emph {et~al.}(2020)\citenamefont
		{Neuh{\"a}user}, \citenamefont {Mellor},\ and\ \citenamefont
		{Lambiotte}}]{neuhauser2019multi}%
	\BibitemOpen
	\bibfield  {author} {\bibinfo {author} {\bibfnamefont {L.}~\bibnamefont
			{Neuh{\"a}user}}, \bibinfo {author} {\bibfnamefont {A.}~\bibnamefont
			{Mellor}}, \ and\ \bibinfo {author} {\bibfnamefont {R.}~\bibnamefont
			{Lambiotte}},\ }\href@noop {} {\bibfield  {journal} {\bibinfo  {journal}
			{Phys. Rev. E}\ }\textbf {\bibinfo {volume} {101}},\ \bibinfo {pages}
		{032310} (\bibinfo {year} {2020})}\BibitemShut {NoStop}%
	\bibitem [{\citenamefont {Alvarez-Rodriguez}\ \emph {et~al.}(2021)\citenamefont
		{Alvarez-Rodriguez}, \citenamefont {Battiston}, \citenamefont {de~Arruda},
		\citenamefont {Moreno}, \citenamefont {Perc},\ and\ \citenamefont
		{Latora}}]{alvarez2021evolutionary}%
	\BibitemOpen
	\bibfield  {author} {\bibinfo {author} {\bibfnamefont {U.}~\bibnamefont
			{Alvarez-Rodriguez}}, \bibinfo {author} {\bibfnamefont {F.}~\bibnamefont
			{Battiston}}, \bibinfo {author} {\bibfnamefont {G.~F.}\ \bibnamefont
			{de~Arruda}}, \bibinfo {author} {\bibfnamefont {Y.}~\bibnamefont {Moreno}},
		\bibinfo {author} {\bibfnamefont {M.}~\bibnamefont {Perc}}, \ and\ \bibinfo
		{author} {\bibfnamefont {V.}~\bibnamefont {Latora}},\ }\href@noop {}
	{\bibfield  {journal} {\bibinfo  {journal} {Nature Human Behaviour}\ }\textbf
		{\bibinfo {volume} {5}},\ \bibinfo {pages} {586} (\bibinfo {year}
		{2021})}\BibitemShut {NoStop}%
	\bibitem [{\citenamefont {Achlioptas}\ \emph {et~al.}(2009)\citenamefont
		{Achlioptas}, \citenamefont {D'Souza},\ and\ \citenamefont
		{Spencer}}]{achlioptas2009explosive}%
	\BibitemOpen
	\bibfield  {author} {\bibinfo {author} {\bibfnamefont {D.}~\bibnamefont
			{Achlioptas}}, \bibinfo {author} {\bibfnamefont {R.~M.}\ \bibnamefont
			{D'Souza}}, \ and\ \bibinfo {author} {\bibfnamefont {J.}~\bibnamefont
			{Spencer}},\ }\href@noop {} {\bibfield  {journal} {\bibinfo  {journal}
			{Science}\ }\textbf {\bibinfo {volume} {323}},\ \bibinfo {pages} {1453}
		(\bibinfo {year} {2009})}\BibitemShut {NoStop}%
	\bibitem [{\citenamefont {da~Costa}\ \emph {et~al.}(2010)\citenamefont
		{da~Costa}, \citenamefont {Dorogovtsev}, \citenamefont {Goltsev},\ and\
		\citenamefont {Mendes}}]{da2010explosive}%
	\BibitemOpen
	\bibfield  {author} {\bibinfo {author} {\bibfnamefont {R.~A.}\ \bibnamefont
			{da~Costa}}, \bibinfo {author} {\bibfnamefont {S.~N.}\ \bibnamefont
			{Dorogovtsev}}, \bibinfo {author} {\bibfnamefont {A.~V.}\ \bibnamefont
			{Goltsev}}, \ and\ \bibinfo {author} {\bibfnamefont {J.~F.~F.}\ \bibnamefont
			{Mendes}},\ }\href@noop {} {\bibfield  {journal} {\bibinfo  {journal} {Phys.
				Rev. Lett.}\ }\textbf {\bibinfo {volume} {105}},\ \bibinfo {pages} {255701}
		(\bibinfo {year} {2010})}\BibitemShut {NoStop}%
	\bibitem [{\citenamefont {Riordan}\ and\ \citenamefont
		{Warnke}(2011)}]{riordan2011explosive}%
	\BibitemOpen
	\bibfield  {author} {\bibinfo {author} {\bibfnamefont {O.}~\bibnamefont
			{Riordan}}\ and\ \bibinfo {author} {\bibfnamefont {L.}~\bibnamefont
			{Warnke}},\ }\href@noop {} {\bibfield  {journal} {\bibinfo  {journal}
			{Science}\ }\textbf {\bibinfo {volume} {333}},\ \bibinfo {pages} {322}
		(\bibinfo {year} {2011})}\BibitemShut {NoStop}%
	\bibitem [{\citenamefont {D'Souza}\ \emph {et~al.}(2019)\citenamefont
		{D'Souza}, \citenamefont {G{\'o}mez-Garde{\~n}es}, \citenamefont {Nagler},\
		and\ \citenamefont {Arenas}}]{d2019explosive}%
	\BibitemOpen
	\bibfield  {author} {\bibinfo {author} {\bibfnamefont {R.~M.}\ \bibnamefont
			{D'Souza}}, \bibinfo {author} {\bibfnamefont {J.}~\bibnamefont
			{G{\'o}mez-Garde{\~n}es}}, \bibinfo {author} {\bibfnamefont {J.}~\bibnamefont
			{Nagler}}, \ and\ \bibinfo {author} {\bibfnamefont {A.}~\bibnamefont
			{Arenas}},\ }\href@noop {} {\bibfield  {journal} {\bibinfo  {journal}
			{Advances in Physics}\ }\textbf {\bibinfo {volume} {68}},\ \bibinfo {pages}
		{123} (\bibinfo {year} {2019})}\BibitemShut {NoStop}%
	\bibitem [{\citenamefont {{G{\'o}mez-Gardenes}}\ \emph
		{et~al.}(2011)\citenamefont {{G{\'o}mez-Gardenes}}, \citenamefont
		{G{\'o}mez}, \citenamefont {Arenas},\ and\ \citenamefont
		{Moreno}}]{gomez2011explosive}%
	\BibitemOpen
	\bibfield  {author} {\bibinfo {author} {\bibfnamefont {J.}~\bibnamefont
			{{G{\'o}mez-Gardenes}}}, \bibinfo {author} {\bibfnamefont {S.}~\bibnamefont
			{G{\'o}mez}}, \bibinfo {author} {\bibfnamefont {A.}~\bibnamefont {Arenas}}, \
		and\ \bibinfo {author} {\bibfnamefont {Y.}~\bibnamefont {Moreno}},\
	}\href@noop {} {\bibfield  {journal} {\bibinfo  {journal} {Phys. Rev. Lett.}\
		}\textbf {\bibinfo {volume} {106}},\ \bibinfo {pages} {128701} (\bibinfo
		{year} {2011})}\BibitemShut {NoStop}%
	\bibitem [{\citenamefont {Matamalas}\ \emph {et~al.}(2020)\citenamefont
		{Matamalas}, \citenamefont {G{\'o}mez},\ and\ \citenamefont
		{Arenas}}]{matamalas2020abrupt}%
	\BibitemOpen
	\bibfield  {author} {\bibinfo {author} {\bibfnamefont {J.~T.}\ \bibnamefont
			{Matamalas}}, \bibinfo {author} {\bibfnamefont {S.}~\bibnamefont
			{G{\'o}mez}}, \ and\ \bibinfo {author} {\bibfnamefont {A.}~\bibnamefont
			{Arenas}},\ }\href@noop {} {\bibfield  {journal} {\bibinfo  {journal} {Phys.
				Rev. Res.}\ }\textbf {\bibinfo {volume} {2}},\ \bibinfo {pages} {012049}
		(\bibinfo {year} {2020})}\BibitemShut {NoStop}%
	\bibitem [{\citenamefont {Chowdhary}\ \emph {et~al.}(2021)\citenamefont
		{Chowdhary}, \citenamefont {Kumar}, \citenamefont {Cencetti}, \citenamefont
		{Iacopini},\ and\ \citenamefont {Battiston}}]{chowdhary2021simplicial}%
	\BibitemOpen
	\bibfield  {author} {\bibinfo {author} {\bibfnamefont {S.}~\bibnamefont
			{Chowdhary}}, \bibinfo {author} {\bibfnamefont {A.}~\bibnamefont {Kumar}},
		\bibinfo {author} {\bibfnamefont {G.}~\bibnamefont {Cencetti}}, \bibinfo
		{author} {\bibfnamefont {I.}~\bibnamefont {Iacopini}}, \ and\ \bibinfo
		{author} {\bibfnamefont {F.}~\bibnamefont {Battiston}},\ }\href@noop {}
	{\bibfield  {journal} {\bibinfo  {journal} {J. Phys.: Complex.}\ } (\bibinfo
		{year} {2021})}\BibitemShut {NoStop}%
	\bibitem [{\citenamefont {St-Onge}\ \emph
		{et~al.}(2021{\natexlab{a}})\citenamefont {St-Onge}, \citenamefont {Sun},
		\citenamefont {Allard}, \citenamefont {H{\'e}bert-Dufresne},\ and\
		\citenamefont {Bianconi}}]{st2021bursty}%
	\BibitemOpen
	\bibfield  {author} {\bibinfo {author} {\bibfnamefont {G.}~\bibnamefont
			{St-Onge}}, \bibinfo {author} {\bibfnamefont {H.}~\bibnamefont {Sun}},
		\bibinfo {author} {\bibfnamefont {A.}~\bibnamefont {Allard}}, \bibinfo
		{author} {\bibfnamefont {L.}~\bibnamefont {H{\'e}bert-Dufresne}}, \ and\
		\bibinfo {author} {\bibfnamefont {G.}~\bibnamefont {Bianconi}},\ }\href@noop
	{} {\bibfield  {journal} {\bibinfo  {journal} {arXiv preprint
				arXiv:2101.07229}\ } (\bibinfo {year} {2021}{\natexlab{a}})}\BibitemShut
	{NoStop}%
	\bibitem [{\citenamefont {Landry}\ and\ \citenamefont
		{Restrepo}(2020)}]{landry2020effect}%
	\BibitemOpen
	\bibfield  {author} {\bibinfo {author} {\bibfnamefont {N.~W.}\ \bibnamefont
			{Landry}}\ and\ \bibinfo {author} {\bibfnamefont {J.~G.}\ \bibnamefont
			{Restrepo}},\ }\href@noop {} {\bibfield  {journal} {\bibinfo  {journal}
			{Chaos: An Interdisciplinary Journal of Nonlinear Science}\ }\textbf
		{\bibinfo {volume} {30}},\ \bibinfo {pages} {103117} (\bibinfo {year}
		{2020})}\BibitemShut {NoStop}%
	\bibitem [{\citenamefont {St-Onge}\ \emph
		{et~al.}(2021{\natexlab{b}})\citenamefont {St-Onge}, \citenamefont
		{Iacopini}, \citenamefont {Latora}, \citenamefont {Barrat}, \citenamefont
		{Petri}, \citenamefont {Allard},\ and\ \citenamefont
		{H{\'e}bert-Dufresne}}]{st2021influential}%
	\BibitemOpen
	\bibfield  {author} {\bibinfo {author} {\bibfnamefont {G.}~\bibnamefont
			{St-Onge}}, \bibinfo {author} {\bibfnamefont {I.}~\bibnamefont {Iacopini}},
		\bibinfo {author} {\bibfnamefont {V.}~\bibnamefont {Latora}}, \bibinfo
		{author} {\bibfnamefont {A.}~\bibnamefont {Barrat}}, \bibinfo {author}
		{\bibfnamefont {G.}~\bibnamefont {Petri}}, \bibinfo {author} {\bibfnamefont
			{A.}~\bibnamefont {Allard}}, \ and\ \bibinfo {author} {\bibfnamefont
			{L.}~\bibnamefont {H{\'e}bert-Dufresne}},\ }\href@noop {} {\bibfield
		{journal} {\bibinfo  {journal} {arXiv preprint arXiv:2105.07092}\ } (\bibinfo
		{year} {2021}{\natexlab{b}})}\BibitemShut {NoStop}%
	\bibitem [{\citenamefont {Sun}\ and\ \citenamefont
		{Bianconi}(2021)}]{sun2021higher}%
	\BibitemOpen
	\bibfield  {author} {\bibinfo {author} {\bibfnamefont {H.}~\bibnamefont
			{Sun}}\ and\ \bibinfo {author} {\bibfnamefont {G.}~\bibnamefont {Bianconi}},\
	}\href@noop {} {\bibfield  {journal} {\bibinfo  {journal} {arXiv preprint
				arXiv:2104.05457}\ } (\bibinfo {year} {2021})}\BibitemShut {NoStop}%
	\bibitem [{\citenamefont {Mill{\'a}n}\ \emph {et~al.}(2020)\citenamefont
		{Mill{\'a}n}, \citenamefont {Torres},\ and\ \citenamefont
		{Bianconi}}]{millan2020explosive}%
	\BibitemOpen
	\bibfield  {author} {\bibinfo {author} {\bibfnamefont {A.~P.}\ \bibnamefont
			{Mill{\'a}n}}, \bibinfo {author} {\bibfnamefont {J.~J.}\ \bibnamefont
			{Torres}}, \ and\ \bibinfo {author} {\bibfnamefont {G.}~\bibnamefont
			{Bianconi}},\ }\href@noop {} {\bibfield  {journal} {\bibinfo  {journal}
			{Physical Review Letters}\ }\textbf {\bibinfo {volume} {124}},\ \bibinfo
		{pages} {218301} (\bibinfo {year} {2020})}\BibitemShut {NoStop}%
	\bibitem [{\citenamefont {Kuramoto}(1975)}]{kuramoto1975self}%
	\BibitemOpen
	\bibfield  {author} {\bibinfo {author} {\bibfnamefont {Y.}~\bibnamefont
			{Kuramoto}},\ }in\ \href@noop {} {\emph {\bibinfo {booktitle} {International
				symposium on mathematical problems in theoretical physics}}}\ (\bibinfo
	{organization} {Springer},\ \bibinfo {year} {1975})\ pp.\ \bibinfo {pages}
	{420--422}\BibitemShut {NoStop}%
	\bibitem [{\citenamefont {Skardal}\ and\ \citenamefont
		{Arenas}(2020)}]{skardal2020higher}%
	\BibitemOpen
	\bibfield  {author} {\bibinfo {author} {\bibfnamefont {P.~S.}\ \bibnamefont
			{Skardal}}\ and\ \bibinfo {author} {\bibfnamefont {A.}~\bibnamefont
			{Arenas}},\ }\href@noop {} {\bibfield  {journal} {\bibinfo  {journal}
			{Commun. Phys.}\ }\textbf {\bibinfo {volume} {3}},\ \bibinfo {pages} {1}
		(\bibinfo {year} {2020})}\BibitemShut {NoStop}%
	\bibitem [{\citenamefont {Lucas}\ \emph {et~al.}(2020)\citenamefont {Lucas},
		\citenamefont {Cencetti},\ and\ \citenamefont
		{Battiston}}]{lucas2020multiorder}%
	\BibitemOpen
	\bibfield  {author} {\bibinfo {author} {\bibfnamefont {M.}~\bibnamefont
			{Lucas}}, \bibinfo {author} {\bibfnamefont {G.}~\bibnamefont {Cencetti}}, \
		and\ \bibinfo {author} {\bibfnamefont {F.}~\bibnamefont {Battiston}},\
	}\href@noop {} {\bibfield  {journal} {\bibinfo  {journal} {Physical Review
				Research}\ }\textbf {\bibinfo {volume} {2}},\ \bibinfo {pages} {033410}
		(\bibinfo {year} {2020})}\BibitemShut {NoStop}%
	\bibitem [{\citenamefont {Gambuzza}\ \emph {et~al.}(2021)\citenamefont
		{Gambuzza}, \citenamefont {Di~Patti}, \citenamefont {Gallo}, \citenamefont
		{Lepri}, \citenamefont {Romance}, \citenamefont {Criado}, \citenamefont
		{Frasca}, \citenamefont {Latora},\ and\ \citenamefont
		{Boccaletti}}]{gambuzza2021stability}%
	\BibitemOpen
	\bibfield  {author} {\bibinfo {author} {\bibfnamefont {L.}~\bibnamefont
			{Gambuzza}}, \bibinfo {author} {\bibfnamefont {F.}~\bibnamefont {Di~Patti}},
		\bibinfo {author} {\bibfnamefont {L.}~\bibnamefont {Gallo}}, \bibinfo
		{author} {\bibfnamefont {S.}~\bibnamefont {Lepri}}, \bibinfo {author}
		{\bibfnamefont {M.}~\bibnamefont {Romance}}, \bibinfo {author} {\bibfnamefont
			{R.}~\bibnamefont {Criado}}, \bibinfo {author} {\bibfnamefont
			{M.}~\bibnamefont {Frasca}}, \bibinfo {author} {\bibfnamefont
			{V.}~\bibnamefont {Latora}}, \ and\ \bibinfo {author} {\bibfnamefont
			{S.}~\bibnamefont {Boccaletti}},\ }\href@noop {} {\bibfield  {journal}
		{\bibinfo  {journal} {Nat. Commun.}\ }\textbf {\bibinfo {volume} {12}},\
		\bibinfo {pages} {1} (\bibinfo {year} {2021})}\BibitemShut {NoStop}%
	\bibitem [{\citenamefont {Zhang}\ \emph {et~al.}(2021)\citenamefont {Zhang},
		\citenamefont {Latora},\ and\ \citenamefont {Motter}}]{zhang2021generalized}%
	\BibitemOpen
	\bibfield  {author} {\bibinfo {author} {\bibfnamefont {Y.}~\bibnamefont
			{Zhang}}, \bibinfo {author} {\bibfnamefont {V.}~\bibnamefont {Latora}}, \
		and\ \bibinfo {author} {\bibfnamefont {A.~E.}\ \bibnamefont {Motter}},\
	}\href@noop {} {\bibfield  {journal} {\bibinfo  {journal} {arXiv preprint
				arXiv:2010.00613}\ } (\bibinfo {year} {2021})}\BibitemShut {NoStop}%
	\bibitem [{\citenamefont {de~Arruda}\ \emph {et~al.}(2021)\citenamefont
		{de~Arruda}, \citenamefont {Tizzani},\ and\ \citenamefont
		{Moreno}}]{de2021phase}%
	\BibitemOpen
	\bibfield  {author} {\bibinfo {author} {\bibfnamefont {G.~F.}\ \bibnamefont
			{de~Arruda}}, \bibinfo {author} {\bibfnamefont {M.}~\bibnamefont {Tizzani}},
		\ and\ \bibinfo {author} {\bibfnamefont {Y.}~\bibnamefont {Moreno}},\
	}\href@noop {} {\bibfield  {journal} {\bibinfo  {journal} {Commun. Phys.}\
		}\textbf {\bibinfo {volume} {4}},\ \bibinfo {pages} {1} (\bibinfo {year}
		{2021})}\BibitemShut {NoStop}%
	\bibitem [{\citenamefont {St-Onge}\ \emph
		{et~al.}(2021{\natexlab{c}})\citenamefont {St-Onge}, \citenamefont
		{Thibeault}, \citenamefont {Allard}, \citenamefont {Dub{\'e}},\ and\
		\citenamefont {H{\'e}bert-Dufresne}}]{stonge2020master}%
	\BibitemOpen
	\bibfield  {author} {\bibinfo {author} {\bibfnamefont {G.}~\bibnamefont
			{St-Onge}}, \bibinfo {author} {\bibfnamefont {V.}~\bibnamefont {Thibeault}},
		\bibinfo {author} {\bibfnamefont {A.}~\bibnamefont {Allard}}, \bibinfo
		{author} {\bibfnamefont {L.~J.}\ \bibnamefont {Dub{\'e}}}, \ and\ \bibinfo
		{author} {\bibfnamefont {L.}~\bibnamefont {H{\'e}bert-Dufresne}},\
	}\href@noop {} {\bibfield  {journal} {\bibinfo  {journal} {Phys. Rev. E}\
		}\textbf {\bibinfo {volume} {103}},\ \bibinfo {pages} {032301} (\bibinfo
		{year} {2021}{\natexlab{c}})}\BibitemShut {NoStop}%
	\bibitem [{\citenamefont {Kuehn}\ and\ \citenamefont
		{Bick}(2021)}]{kuehn2021universal}%
	\BibitemOpen
	\bibfield  {author} {\bibinfo {author} {\bibfnamefont {C.}~\bibnamefont
			{Kuehn}}\ and\ \bibinfo {author} {\bibfnamefont {C.}~\bibnamefont {Bick}},\
	}\href@noop {} {\bibfield  {journal} {\bibinfo  {journal} {Sci. Adv.}\
		}\textbf {\bibinfo {volume} {7}},\ \bibinfo {pages} {eabe3824} (\bibinfo
		{year} {2021})}\BibitemShut {NoStop}%
	\bibitem [{\citenamefont {Cisneros-Velarde}\ and\ \citenamefont
		{Bullo}(2020)}]{cisneros2020multi}%
	\BibitemOpen
	\bibfield  {author} {\bibinfo {author} {\bibfnamefont {P.}~\bibnamefont
			{Cisneros-Velarde}}\ and\ \bibinfo {author} {\bibfnamefont {F.}~\bibnamefont
			{Bullo}},\ }\href@noop {} {\bibfield  {journal} {\bibinfo  {journal}
			{arXiv:2005.11404}\ } (\bibinfo {year} {2020})}\BibitemShut {NoStop}%
	\bibitem [{\citenamefont {Porter}\ and\ \citenamefont
		{Gleeson}(2016)}]{porter2016dynamical}%
	\BibitemOpen
	\bibfield  {author} {\bibinfo {author} {\bibfnamefont {M.}~\bibnamefont
			{Porter}}\ and\ \bibinfo {author} {\bibfnamefont {J.}~\bibnamefont
			{Gleeson}},\ }\href@noop {} {\emph {\bibinfo {title} {Dynamical Systems on
				Networks: A Tutorial}}},\ Frontiers in Applied Dynamical Systems: Reviews and
	Tutorials\ (\bibinfo  {publisher} {Springer International Publishing},\
	\bibinfo {year} {2016})\BibitemShut {NoStop}%
	\bibitem [{\citenamefont {Ghorbanchian}\ \emph {et~al.}(2021)\citenamefont
		{Ghorbanchian}, \citenamefont {Restrepo}, \citenamefont {Torres},\ and\
		\citenamefont {Bianconi}}]{ghorbanchian2020higher}%
	\BibitemOpen
	\bibfield  {author} {\bibinfo {author} {\bibfnamefont {R.}~\bibnamefont
			{Ghorbanchian}}, \bibinfo {author} {\bibfnamefont {J.~G.}\ \bibnamefont
			{Restrepo}}, \bibinfo {author} {\bibfnamefont {J.~J.}\ \bibnamefont
			{Torres}}, \ and\ \bibinfo {author} {\bibfnamefont {G.}~\bibnamefont
			{Bianconi}},\ }\href@noop {} {\bibfield  {journal} {\bibinfo  {journal}
			{Commun. Phys.}\ }\textbf {\bibinfo {volume} {4}},\ \bibinfo {pages} {1}
		(\bibinfo {year} {2021})}\BibitemShut {NoStop}%
	\bibitem [{\citenamefont {Schaub}\ and\ \citenamefont
		{Segarra}(2018)}]{schaub2018flow}%
	\BibitemOpen
	\bibfield  {author} {\bibinfo {author} {\bibfnamefont {M.~T.}\ \bibnamefont
			{Schaub}}\ and\ \bibinfo {author} {\bibfnamefont {S.}~\bibnamefont
			{Segarra}},\ }in\ \href@noop {} {\emph {\bibinfo {booktitle} {2018 {{IEEE}}
				Global Conference on Signal and Information Processing ({{GlobalSIP}})}}}\
	(\bibinfo {organization} {{IEEE}},\ \bibinfo {year} {2018})\ pp.\ \bibinfo
	{pages} {735--739}\BibitemShut {NoStop}%
	\bibitem [{\citenamefont {Barbarossa}\ and\ \citenamefont
		{Sardellitti}(2020)}]{barbarossa2020topological}%
	\BibitemOpen
	\bibfield  {author} {\bibinfo {author} {\bibfnamefont {S.}~\bibnamefont
			{Barbarossa}}\ and\ \bibinfo {author} {\bibfnamefont {S.}~\bibnamefont
			{Sardellitti}},\ }\href@noop {} {\bibfield  {journal} {\bibinfo  {journal}
			{IEEE Transactions on Signal Processing}\ } (\bibinfo {year}
		{2020})}\BibitemShut {NoStop}%
	\bibitem [{\citenamefont {Holme}\ and\ \citenamefont
		{Saram{\"a}ki}(2012)}]{holme2012temporal}%
	\BibitemOpen
	\bibfield  {author} {\bibinfo {author} {\bibfnamefont {P.}~\bibnamefont
			{Holme}}\ and\ \bibinfo {author} {\bibfnamefont {J.}~\bibnamefont
			{Saram{\"a}ki}},\ }\href@noop {} {\bibfield  {journal} {\bibinfo  {journal}
			{Phys. Rep.}\ }\textbf {\bibinfo {volume} {519}},\ \bibinfo {pages} {97}
		(\bibinfo {year} {2012})}\BibitemShut {NoStop}%
	\bibitem [{\citenamefont {Gross}\ and\ \citenamefont
		{Blasius}(2008)}]{gross2008adaptive}%
	\BibitemOpen
	\bibfield  {author} {\bibinfo {author} {\bibfnamefont {T.}~\bibnamefont
			{Gross}}\ and\ \bibinfo {author} {\bibfnamefont {B.}~\bibnamefont
			{Blasius}},\ }\href@noop {} {\bibfield  {journal} {\bibinfo  {journal}
			{Journal of the Royal Society Interface}\ }\textbf {\bibinfo {volume} {5}},\
		\bibinfo {pages} {259} (\bibinfo {year} {2008})}\BibitemShut {NoStop}%
	\bibitem [{\citenamefont {Young}\ \emph {et~al.}(2021)\citenamefont {Young},
		\citenamefont {Petri},\ and\ \citenamefont {Peixoto}}]{young2020hypergraph}%
	\BibitemOpen
	\bibfield  {author} {\bibinfo {author} {\bibfnamefont {J.-G.}\ \bibnamefont
			{Young}}, \bibinfo {author} {\bibfnamefont {G.}~\bibnamefont {Petri}}, \ and\
		\bibinfo {author} {\bibfnamefont {T.~P.}\ \bibnamefont {Peixoto}},\
	}\href@noop {} {\bibfield  {journal} {\bibinfo  {journal} {Commun. Phys.}\
		}\textbf {\bibinfo {volume} {4}},\ \bibinfo {pages} {1} (\bibinfo {year}
		{2021})}\BibitemShut {NoStop}%
	\bibitem [{\citenamefont {Saebi}\ \emph {et~al.}(2020)\citenamefont {Saebi},
		\citenamefont {Ciampaglia}, \citenamefont {Kaplan},\ and\ \citenamefont
		{Chawla}}]{saebi2020honem}%
	\BibitemOpen
	\bibfield  {author} {\bibinfo {author} {\bibfnamefont {M.}~\bibnamefont
			{Saebi}}, \bibinfo {author} {\bibfnamefont {G.~L.}\ \bibnamefont
			{Ciampaglia}}, \bibinfo {author} {\bibfnamefont {L.~M.}\ \bibnamefont
			{Kaplan}}, \ and\ \bibinfo {author} {\bibfnamefont {N.~V.}\ \bibnamefont
			{Chawla}},\ }\href@noop {} {\bibfield  {journal} {\bibinfo  {journal} {Big
				Data}\ }\textbf {\bibinfo {volume} {8}},\ \bibinfo {pages} {255} (\bibinfo
		{year} {2020})}\BibitemShut {NoStop}%
	\bibitem [{\citenamefont {Newman}(2018)}]{newman2018structure}%
	\BibitemOpen
	\bibfield  {author} {\bibinfo {author} {\bibfnamefont {M.~E.~J.}\
			\bibnamefont {Newman}},\ }\href@noop {} {\bibfield  {journal} {\bibinfo
			{journal} {Nature Physics}\ ,\ \bibinfo {pages} {542}} (\bibinfo {year}
		{2018})}\BibitemShut {NoStop}%
	\bibitem [{\citenamefont {Peixoto}(2018)}]{peixoto2018network}%
	\BibitemOpen
	\bibfield  {author} {\bibinfo {author} {\bibfnamefont {T.~P.}\ \bibnamefont
			{Peixoto}},\ }\href@noop {} {\bibfield  {journal} {\bibinfo  {journal} {Phys.
				Rev. X}\ ,\ \bibinfo {pages} {041011}} (\bibinfo {year} {2018})}\BibitemShut
	{NoStop}%
	\bibitem [{\citenamefont {Musciotto}\ \emph {et~al.}(2021)\citenamefont
		{Musciotto}, \citenamefont {Battiston},\ and\ \citenamefont
		{Mantegna}}]{musciotto2021detecting}%
	\BibitemOpen
	\bibfield  {author} {\bibinfo {author} {\bibfnamefont {F.}~\bibnamefont
			{Musciotto}}, \bibinfo {author} {\bibfnamefont {F.}~\bibnamefont
			{Battiston}}, \ and\ \bibinfo {author} {\bibfnamefont {R.~N.}\ \bibnamefont
			{Mantegna}},\ }\href@noop {} {\bibfield  {journal} {\bibinfo  {journal}
			{arXiv preprint arXiv:2103.16484}\ } (\bibinfo {year} {2021})}\BibitemShut
	{NoStop}%
	\bibitem [{\citenamefont {Bohland}\ \emph {et~al.}(2009)\citenamefont
		{Bohland}, \citenamefont {Wu}, \citenamefont {Barbas}, \citenamefont {Bokil},
		\citenamefont {Bota}, \citenamefont {Breiter}, \citenamefont {Cline},
		\citenamefont {Doyle}, \citenamefont {Freed}, \citenamefont {Greenspan} \emph
		{et~al.}}]{bohland2009proposal}%
	\BibitemOpen
	\bibfield  {author} {\bibinfo {author} {\bibfnamefont {J.~W.}\ \bibnamefont
			{Bohland}}, \bibinfo {author} {\bibfnamefont {C.}~\bibnamefont {Wu}},
		\bibinfo {author} {\bibfnamefont {H.}~\bibnamefont {Barbas}}, \bibinfo
		{author} {\bibfnamefont {H.}~\bibnamefont {Bokil}}, \bibinfo {author}
		{\bibfnamefont {M.}~\bibnamefont {Bota}}, \bibinfo {author} {\bibfnamefont
			{H.~C.}\ \bibnamefont {Breiter}}, \bibinfo {author} {\bibfnamefont {H.~T.}\
			\bibnamefont {Cline}}, \bibinfo {author} {\bibfnamefont {J.~C.}\ \bibnamefont
			{Doyle}}, \bibinfo {author} {\bibfnamefont {P.~J.}\ \bibnamefont {Freed}},
		\bibinfo {author} {\bibfnamefont {R.~J.}\ \bibnamefont {Greenspan}},  \emph
		{et~al.},\ }\href@noop {} {\bibfield  {journal} {\bibinfo  {journal} {PLoS
				Comput Biol}\ }\textbf {\bibinfo {volume} {5}},\ \bibinfo {pages} {e1000334}
		(\bibinfo {year} {2009})}\BibitemShut {NoStop}%
	\bibitem [{\citenamefont {Bassett}\ and\ \citenamefont
		{Sporns}(2017)}]{bassett2017network}%
	\BibitemOpen
	\bibfield  {author} {\bibinfo {author} {\bibfnamefont {D.~S.}\ \bibnamefont
			{Bassett}}\ and\ \bibinfo {author} {\bibfnamefont {O.}~\bibnamefont
			{Sporns}},\ }\href@noop {} {\bibfield  {journal} {\bibinfo  {journal} {Nat.
				Neurosci.}\ }\textbf {\bibinfo {volume} {20}},\ \bibinfo {pages} {353}
		(\bibinfo {year} {2017})}\BibitemShut {NoStop}%
	\bibitem [{\citenamefont {Glomb}\ \emph {et~al.}(2021)\citenamefont {Glomb},
		\citenamefont {Cabral}, \citenamefont {Cattani}, \citenamefont {Mazzoni},
		\citenamefont {Raj},\ and\ \citenamefont
		{Franceschiello}}]{glomb2021computational}%
	\BibitemOpen
	\bibfield  {author} {\bibinfo {author} {\bibfnamefont {K.}~\bibnamefont
			{Glomb}}, \bibinfo {author} {\bibfnamefont {J.}~\bibnamefont {Cabral}},
		\bibinfo {author} {\bibfnamefont {A.}~\bibnamefont {Cattani}}, \bibinfo
		{author} {\bibfnamefont {A.}~\bibnamefont {Mazzoni}}, \bibinfo {author}
		{\bibfnamefont {A.}~\bibnamefont {Raj}}, \ and\ \bibinfo {author}
		{\bibfnamefont {B.}~\bibnamefont {Franceschiello}},\ }\href@noop {}
	{\bibfield  {journal} {\bibinfo  {journal} {Brain Topography}\ ,\ \bibinfo
			{pages} {1}} (\bibinfo {year} {2021})}\BibitemShut {NoStop}%
	\bibitem [{\citenamefont {Skudlarski}\ \emph {et~al.}(2008)\citenamefont
		{Skudlarski}, \citenamefont {Jagannathan}, \citenamefont {Calhoun},
		\citenamefont {Hampson}, \citenamefont {Skudlarska},\ and\ \citenamefont
		{Pearlson}}]{skudlarski2008measuring}%
	\BibitemOpen
	\bibfield  {author} {\bibinfo {author} {\bibfnamefont {P.}~\bibnamefont
			{Skudlarski}}, \bibinfo {author} {\bibfnamefont {K.}~\bibnamefont
			{Jagannathan}}, \bibinfo {author} {\bibfnamefont {V.~D.}\ \bibnamefont
			{Calhoun}}, \bibinfo {author} {\bibfnamefont {M.}~\bibnamefont {Hampson}},
		\bibinfo {author} {\bibfnamefont {B.~A.}\ \bibnamefont {Skudlarska}}, \ and\
		\bibinfo {author} {\bibfnamefont {G.}~\bibnamefont {Pearlson}},\ }\href@noop
	{} {\bibfield  {journal} {\bibinfo  {journal} {NeuroImage}\ }\textbf
		{\bibinfo {volume} {43}},\ \bibinfo {pages} {554?561} (\bibinfo {year}
		{2008})}\BibitemShut {NoStop}%
	\bibitem [{\citenamefont {Tass}\ \emph {et~al.}(1998)\citenamefont {Tass},
		\citenamefont {Rosenblum}, \citenamefont {Weule}, \citenamefont {Kurths},
		\citenamefont {Pikovsky}, \citenamefont {Volkmann}, \citenamefont
		{Schnitzler},\ and\ \citenamefont {Freund}}]{tass1998detection}%
	\BibitemOpen
	\bibfield  {author} {\bibinfo {author} {\bibfnamefont {P.}~\bibnamefont
			{Tass}}, \bibinfo {author} {\bibfnamefont {M.~G.}\ \bibnamefont {Rosenblum}},
		\bibinfo {author} {\bibfnamefont {J.}~\bibnamefont {Weule}}, \bibinfo
		{author} {\bibfnamefont {J.}~\bibnamefont {Kurths}}, \bibinfo {author}
		{\bibfnamefont {A.}~\bibnamefont {Pikovsky}}, \bibinfo {author}
		{\bibfnamefont {J.}~\bibnamefont {Volkmann}}, \bibinfo {author}
		{\bibfnamefont {A.}~\bibnamefont {Schnitzler}}, \ and\ \bibinfo {author}
		{\bibfnamefont {H.-J.}\ \bibnamefont {Freund}},\ }\href@noop {} {\bibfield
		{journal} {\bibinfo  {journal} {Phys. Rev. Lett.}\ }\textbf {\bibinfo
			{volume} {81}},\ \bibinfo {pages} {3291} (\bibinfo {year}
		{1998})}\BibitemShut {NoStop}%
	\bibitem [{\citenamefont {Kralemann}\ \emph {et~al.}(2008)\citenamefont
		{Kralemann}, \citenamefont {Cimponeriu}, \citenamefont {Rosenblum},
		\citenamefont {Pikovsky},\ and\ \citenamefont {Mrowka}}]{kralemann2008phase}%
	\BibitemOpen
	\bibfield  {author} {\bibinfo {author} {\bibfnamefont {B.}~\bibnamefont
			{Kralemann}}, \bibinfo {author} {\bibfnamefont {L.}~\bibnamefont
			{Cimponeriu}}, \bibinfo {author} {\bibfnamefont {M.}~\bibnamefont
			{Rosenblum}}, \bibinfo {author} {\bibfnamefont {A.}~\bibnamefont {Pikovsky}},
		\ and\ \bibinfo {author} {\bibfnamefont {R.}~\bibnamefont {Mrowka}},\
	}\href@noop {} {\bibfield  {journal} {\bibinfo  {journal} {Phys. Rev. E}\
		}\textbf {\bibinfo {volume} {77}},\ \bibinfo {pages} {066205} (\bibinfo
		{year} {2008})}\BibitemShut {NoStop}%
	\bibitem [{\citenamefont {Kralemann}\ \emph {et~al.}(2011)\citenamefont
		{Kralemann}, \citenamefont {Pikovsky},\ and\ \citenamefont
		{Rosenblum}}]{kralemann2011reconstructing}%
	\BibitemOpen
	\bibfield  {author} {\bibinfo {author} {\bibfnamefont {B.}~\bibnamefont
			{Kralemann}}, \bibinfo {author} {\bibfnamefont {A.}~\bibnamefont {Pikovsky}},
		\ and\ \bibinfo {author} {\bibfnamefont {M.}~\bibnamefont {Rosenblum}},\
	}\href@noop {} {\bibfield  {journal} {\bibinfo  {journal} {Chaos}\ }\textbf
		{\bibinfo {volume} {21}},\ \bibinfo {pages} {025104} (\bibinfo {year}
		{2011})}\BibitemShut {NoStop}%
	\bibitem [{\citenamefont {Kralemann}\ \emph {et~al.}(2013)\citenamefont
		{Kralemann}, \citenamefont {Pikovsky},\ and\ \citenamefont
		{Rosenblum}}]{kralemann2013detecting}%
	\BibitemOpen
	\bibfield  {author} {\bibinfo {author} {\bibfnamefont {B.}~\bibnamefont
			{Kralemann}}, \bibinfo {author} {\bibfnamefont {A.}~\bibnamefont {Pikovsky}},
		\ and\ \bibinfo {author} {\bibfnamefont {M.}~\bibnamefont {Rosenblum}},\
	}\href@noop {} {\bibfield  {journal} {\bibinfo  {journal} {Phys. Rev. E}\
		}\textbf {\bibinfo {volume} {87}},\ \bibinfo {pages} {052904} (\bibinfo
		{year} {2013})}\BibitemShut {NoStop}%
	\bibitem [{\citenamefont {Kralemann}\ \emph {et~al.}(2014)\citenamefont
		{Kralemann}, \citenamefont {Pikovsky},\ and\ \citenamefont
		{Rosenblum}}]{kralemann2014reconstructing}%
	\BibitemOpen
	\bibfield  {author} {\bibinfo {author} {\bibfnamefont {B.}~\bibnamefont
			{Kralemann}}, \bibinfo {author} {\bibfnamefont {A.}~\bibnamefont {Pikovsky}},
		\ and\ \bibinfo {author} {\bibfnamefont {M.}~\bibnamefont {Rosenblum}},\
	}\href@noop {} {\bibfield  {journal} {\bibinfo  {journal} {New J. Phys.}\
		}\textbf {\bibinfo {volume} {16}},\ \bibinfo {pages} {085013} (\bibinfo
		{year} {2014})}\BibitemShut {NoStop}%
	\bibitem [{\citenamefont {Pikovsky}(2018)}]{pikovsky2018reconstruction}%
	\BibitemOpen
	\bibfield  {author} {\bibinfo {author} {\bibfnamefont {A.}~\bibnamefont
			{Pikovsky}},\ }\href@noop {} {\bibfield  {journal} {\bibinfo  {journal}
			{Physics Letters A}\ }\textbf {\bibinfo {volume} {382}},\ \bibinfo {pages}
		{147} (\bibinfo {year} {2018})}\BibitemShut {NoStop}%
	\bibitem [{\citenamefont {Barnett}\ \emph {et~al.}(2009)\citenamefont
		{Barnett}, \citenamefont {Barrett},\ and\ \citenamefont
		{Seth}}]{barnett2009granger}%
	\BibitemOpen
	\bibfield  {author} {\bibinfo {author} {\bibfnamefont {L.}~\bibnamefont
			{Barnett}}, \bibinfo {author} {\bibfnamefont {A.~B.}\ \bibnamefont
			{Barrett}}, \ and\ \bibinfo {author} {\bibfnamefont {A.~K.}\ \bibnamefont
			{Seth}},\ }\href@noop {} {\bibfield  {journal} {\bibinfo  {journal} {Phys.
				Rev. Lett.}\ }\textbf {\bibinfo {volume} {103}},\ \bibinfo {pages} {238701}
		(\bibinfo {year} {2009})}\BibitemShut {NoStop}%
	\bibitem [{\citenamefont {Schreiber}(2000)}]{schreiber2000measuring}%
	\BibitemOpen
	\bibfield  {author} {\bibinfo {author} {\bibfnamefont {T.}~\bibnamefont
			{Schreiber}},\ }\href@noop {} {\bibfield  {journal} {\bibinfo  {journal}
			{Phys. Rev. Lett.}\ }\textbf {\bibinfo {volume} {85}},\ \bibinfo {pages}
		{461} (\bibinfo {year} {2000})}\BibitemShut {NoStop}%
	\bibitem [{\citenamefont {Rosas}\ \emph {et~al.}(2019)\citenamefont {Rosas},
		\citenamefont {Mediano}, \citenamefont {Gastpar},\ and\ \citenamefont
		{Jensen}}]{rosas2019quantifying}%
	\BibitemOpen
	\bibfield  {author} {\bibinfo {author} {\bibfnamefont {F.~E.}\ \bibnamefont
			{Rosas}}, \bibinfo {author} {\bibfnamefont {P.~A.}\ \bibnamefont {Mediano}},
		\bibinfo {author} {\bibfnamefont {M.}~\bibnamefont {Gastpar}}, \ and\
		\bibinfo {author} {\bibfnamefont {H.~J.}\ \bibnamefont {Jensen}},\
	}\href@noop {} {\bibfield  {journal} {\bibinfo  {journal} {Phys. Rev. E}\
		}\textbf {\bibinfo {volume} {100}},\ \bibinfo {pages} {032305} (\bibinfo
		{year} {2019})}\BibitemShut {NoStop}%
	\bibitem [{\citenamefont {Pearl}(2009)}]{pearl2009causality}%
	\BibitemOpen
	\bibfield  {author} {\bibinfo {author} {\bibfnamefont {J.}~\bibnamefont
			{Pearl}},\ }\href@noop {} {\emph {\bibinfo {title} {Causality}}}\ (\bibinfo
	{publisher} {Cambridge university press},\ \bibinfo {year}
	{2009})\BibitemShut {NoStop}%
	\bibitem [{\citenamefont {Peixoto}(2019)}]{peixoto2019network}%
	\BibitemOpen
	\bibfield  {author} {\bibinfo {author} {\bibfnamefont {T.~P.}\ \bibnamefont
			{Peixoto}},\ }\href@noop {} {\bibfield  {journal} {\bibinfo  {journal} {Phys.
				Rev. Lett.}\ }\textbf {\bibinfo {volume} {123}},\ \bibinfo {pages} {128301}
		(\bibinfo {year} {2019})}\BibitemShut {NoStop}%
	\bibitem [{\citenamefont {Bianconi}\ \emph {et~al.}(2014)\citenamefont
		{Bianconi}, \citenamefont {Darst}, \citenamefont {Iacovacci},\ and\
		\citenamefont {Fortunato}}]{bianconi2014triadic}%
	\BibitemOpen
	\bibfield  {author} {\bibinfo {author} {\bibfnamefont {G.}~\bibnamefont
			{Bianconi}}, \bibinfo {author} {\bibfnamefont {R.~K.}\ \bibnamefont {Darst}},
		\bibinfo {author} {\bibfnamefont {J.}~\bibnamefont {Iacovacci}}, \ and\
		\bibinfo {author} {\bibfnamefont {S.}~\bibnamefont {Fortunato}},\ }\href@noop
	{} {\bibfield  {journal} {\bibinfo  {journal} {Phys. Rev. E}\ }\textbf
		{\bibinfo {volume} {90}},\ \bibinfo {pages} {042806} (\bibinfo {year}
		{2014})}\BibitemShut {NoStop}%
	\bibitem [{\citenamefont {Courtney}\ and\ \citenamefont
		{Bianconi}(2016)}]{courtney2016generalized}%
	\BibitemOpen
	\bibfield  {author} {\bibinfo {author} {\bibfnamefont {O.~T.}\ \bibnamefont
			{Courtney}}\ and\ \bibinfo {author} {\bibfnamefont {G.}~\bibnamefont
			{Bianconi}},\ }\href@noop {} {\bibfield  {journal} {\bibinfo  {journal}
			{Phys. Rev. E}\ }\textbf {\bibinfo {volume} {93}},\ \bibinfo {pages} {062311}
		(\bibinfo {year} {2016})}\BibitemShut {NoStop}%
	\bibitem [{\citenamefont {Chodrow}(2020)}]{chodrow2020configuration}%
	\BibitemOpen
	\bibfield  {author} {\bibinfo {author} {\bibfnamefont {P.~S.}\ \bibnamefont
			{Chodrow}},\ }\href@noop {} {\bibfield  {journal} {\bibinfo  {journal} {J.
				Complex Netw.}\ }\textbf {\bibinfo {volume} {8}} (\bibinfo {year}
		{2020})}\BibitemShut {NoStop}%
	\bibitem [{\citenamefont {Chodrow}\ \emph {et~al.}(2021)\citenamefont
		{Chodrow}, \citenamefont {Veldt},\ and\ \citenamefont
		{Benson}}]{chodrow2021generative}%
	\BibitemOpen
	\bibfield  {author} {\bibinfo {author} {\bibfnamefont {P.~S.}\ \bibnamefont
			{Chodrow}}, \bibinfo {author} {\bibfnamefont {N.}~\bibnamefont {Veldt}}, \
		and\ \bibinfo {author} {\bibfnamefont {A.~R.}\ \bibnamefont {Benson}},\
	}\href@noop {} {\bibfield  {journal} {\bibinfo  {journal} {Sci. Adv.}\
		}\textbf {\bibinfo {volume} {7}},\ \bibinfo {pages} {eabh1303} (\bibinfo
		{year} {2021})},\ \bibinfo {note} {publisher: American Association for the
		Advancement of Science Section: Research Article}\BibitemShut {NoStop}%
	\bibitem [{\citenamefont {Courtney}\ and\ \citenamefont
		{Bianconi}(2017)}]{courtney2017weighted}%
	\BibitemOpen
	\bibfield  {author} {\bibinfo {author} {\bibfnamefont {O.~T.}\ \bibnamefont
			{Courtney}}\ and\ \bibinfo {author} {\bibfnamefont {G.}~\bibnamefont
			{Bianconi}},\ }\href@noop {} {\bibfield  {journal} {\bibinfo  {journal}
			{Phys. Rev. E}\ }\textbf {\bibinfo {volume} {95}},\ \bibinfo {pages} {062301}
		(\bibinfo {year} {2017})}\BibitemShut {NoStop}%
	\bibitem [{\citenamefont {Kovalenko}\ \emph {et~al.}(2021)\citenamefont
		{Kovalenko}, \citenamefont {Sendi{\~n}a-Nadal}, \citenamefont {Khalil},
		\citenamefont {Dainiak}, \citenamefont {Musatov}, \citenamefont
		{Raigorodskii}, \citenamefont {Alfaro-Bittner}, \citenamefont {Barzel},\ and\
		\citenamefont {Boccaletti}}]{kovalenko2021growing}%
	\BibitemOpen
	\bibfield  {author} {\bibinfo {author} {\bibfnamefont {K.}~\bibnamefont
			{Kovalenko}}, \bibinfo {author} {\bibfnamefont {I.}~\bibnamefont
			{Sendi{\~n}a-Nadal}}, \bibinfo {author} {\bibfnamefont {N.}~\bibnamefont
			{Khalil}}, \bibinfo {author} {\bibfnamefont {A.}~\bibnamefont {Dainiak}},
		\bibinfo {author} {\bibfnamefont {D.}~\bibnamefont {Musatov}}, \bibinfo
		{author} {\bibfnamefont {A.~M.}\ \bibnamefont {Raigorodskii}}, \bibinfo
		{author} {\bibfnamefont {K.}~\bibnamefont {Alfaro-Bittner}}, \bibinfo
		{author} {\bibfnamefont {B.}~\bibnamefont {Barzel}}, \ and\ \bibinfo {author}
		{\bibfnamefont {S.}~\bibnamefont {Boccaletti}},\ }\href@noop {} {\bibfield
		{journal} {\bibinfo  {journal} {Communications Physics}\ }\textbf {\bibinfo
			{volume} {4}},\ \bibinfo {pages} {1} (\bibinfo {year} {2021})}\BibitemShut
	{NoStop}%
	\bibitem [{\citenamefont {Peixoto}(2021)}]{peixoto2021disentangling}%
	\BibitemOpen
	\bibfield  {author} {\bibinfo {author} {\bibfnamefont {T.~P.}\ \bibnamefont
			{Peixoto}},\ }\href@noop {} {\bibfield  {journal} {\bibinfo  {journal}
			{arXiv:2101.02510 [physics, stat]}\ } (\bibinfo {year} {2021})},\ \bibinfo
	{note} {arXiv: 2101.02510}\BibitemShut {NoStop}%
	\bibitem [{\citenamefont {Cencetti}\ \emph {et~al.}(2021)\citenamefont
		{Cencetti}, \citenamefont {Battiston}, \citenamefont {Lepri},\ and\
		\citenamefont {Karsai}}]{cencetti2021temporal}%
	\BibitemOpen
	\bibfield  {author} {\bibinfo {author} {\bibfnamefont {G.}~\bibnamefont
			{Cencetti}}, \bibinfo {author} {\bibfnamefont {F.}~\bibnamefont {Battiston}},
		\bibinfo {author} {\bibfnamefont {B.}~\bibnamefont {Lepri}}, \ and\ \bibinfo
		{author} {\bibfnamefont {M.}~\bibnamefont {Karsai}},\ }\href@noop {}
	{\bibfield  {journal} {\bibinfo  {journal} {Scientific reports}\ }\textbf
		{\bibinfo {volume} {11}},\ \bibinfo {pages} {1} (\bibinfo {year}
		{2021})}\BibitemShut {NoStop}%
	\bibitem [{\citenamefont {Bianconi}\ and\ \citenamefont
		{Rahmede}(2017)}]{bianconi2017emergent}%
	\BibitemOpen
	\bibfield  {author} {\bibinfo {author} {\bibfnamefont {G.}~\bibnamefont
			{Bianconi}}\ and\ \bibinfo {author} {\bibfnamefont {C.}~\bibnamefont
			{Rahmede}},\ }\href@noop {} {\bibfield  {journal} {\bibinfo  {journal} {Sci.
				Rep.}\ }\textbf {\bibinfo {volume} {7}},\ \bibinfo {pages} {1} (\bibinfo
		{year} {2017})}\BibitemShut {NoStop}%
	\bibitem [{\citenamefont {Torres}\ \emph {et~al.}(2020)\citenamefont {Torres},
		\citenamefont {Blevins}, \citenamefont {Bassett},\ and\ \citenamefont
		{Eliassi-Rad}}]{torres2020representations}%
	\BibitemOpen
	\bibfield  {author} {\bibinfo {author} {\bibfnamefont {L.}~\bibnamefont
			{Torres}}, \bibinfo {author} {\bibfnamefont {A.~S.}\ \bibnamefont {Blevins}},
		\bibinfo {author} {\bibfnamefont {D.~S.}\ \bibnamefont {Bassett}}, \ and\
		\bibinfo {author} {\bibfnamefont {T.}~\bibnamefont {Eliassi-Rad}},\
	}\href@noop {} {\bibfield  {journal} {\bibinfo  {journal} {arXiv:2006.02870}\
		} (\bibinfo {year} {2020})}\BibitemShut {NoStop}%
	\bibitem [{\citenamefont {Bick}\ \emph {et~al.}(2021)\citenamefont {Bick},
		\citenamefont {Gross}, \citenamefont {Harrington},\ and\ \citenamefont
		{Schaub}}]{bick2021higher}%
	\BibitemOpen
	\bibfield  {author} {\bibinfo {author} {\bibfnamefont {C.}~\bibnamefont
			{Bick}}, \bibinfo {author} {\bibfnamefont {E.}~\bibnamefont {Gross}},
		\bibinfo {author} {\bibfnamefont {H.~A.}\ \bibnamefont {Harrington}}, \ and\
		\bibinfo {author} {\bibfnamefont {M.~T.}\ \bibnamefont {Schaub}},\
	}\href@noop {} {\bibfield  {journal} {\bibinfo  {journal} {arXiv:2104.11329}\
		} (\bibinfo {year} {2021})}\BibitemShut {NoStop}%
	\bibitem [{\citenamefont {Benson}\ \emph {et~al.}(2021)\citenamefont {Benson},
		\citenamefont {Gleich},\ and\ \citenamefont {Higham}}]{benson2021higher}%
	\BibitemOpen
	\bibfield  {author} {\bibinfo {author} {\bibfnamefont {A.~R.}\ \bibnamefont
			{Benson}}, \bibinfo {author} {\bibfnamefont {D.~F.}\ \bibnamefont {Gleich}},
		\ and\ \bibinfo {author} {\bibfnamefont {D.~J.}\ \bibnamefont {Higham}},\
	}\href@noop {} {\bibfield  {journal} {\bibinfo  {journal} {arXiv preprint
				arXiv:2103.05031}\ } (\bibinfo {year} {2021})}\BibitemShut {NoStop}%
	\bibitem [{\citenamefont {Golubitsky}\ and\ \citenamefont
		{Stewart}(2006)}]{golubitsky2006nonlinear}%
	\BibitemOpen
	\bibfield  {author} {\bibinfo {author} {\bibfnamefont {M.}~\bibnamefont
			{Golubitsky}}\ and\ \bibinfo {author} {\bibfnamefont {I.}~\bibnamefont
			{Stewart}},\ }\href@noop {} {\bibfield  {journal} {\bibinfo  {journal} {Bull.
				Am. Math. Soc}\ }\textbf {\bibinfo {volume} {43}},\ \bibinfo {pages} {305}
		(\bibinfo {year} {2006})}\BibitemShut {NoStop}%
	\bibitem [{\citenamefont {Stewart}\ \emph {et~al.}(2003)\citenamefont
		{Stewart}, \citenamefont {Golubitsky},\ and\ \citenamefont
		{Pivato}}]{stewart2003symmetry}%
	\BibitemOpen
	\bibfield  {author} {\bibinfo {author} {\bibfnamefont {I.}~\bibnamefont
			{Stewart}}, \bibinfo {author} {\bibfnamefont {M.}~\bibnamefont {Golubitsky}},
		\ and\ \bibinfo {author} {\bibfnamefont {M.}~\bibnamefont {Pivato}},\
	}\href@noop {} {\bibfield  {journal} {\bibinfo  {journal} {SIAM Journal on
				Applied Dynamical Systems}\ }\textbf {\bibinfo {volume} {2}},\ \bibinfo
		{pages} {609} (\bibinfo {year} {2003})}\BibitemShut {NoStop}%
	\bibitem [{\citenamefont {Golubitsky}\ \emph {et~al.}(2005)\citenamefont
		{Golubitsky}, \citenamefont {Stewart},\ and\ \citenamefont
		{T{\"o}r{\"o}k}}]{golubitsky2005patterns}%
	\BibitemOpen
	\bibfield  {author} {\bibinfo {author} {\bibfnamefont {M.}~\bibnamefont
			{Golubitsky}}, \bibinfo {author} {\bibfnamefont {I.}~\bibnamefont {Stewart}},
		\ and\ \bibinfo {author} {\bibfnamefont {A.}~\bibnamefont {T{\"o}r{\"o}k}},\
	}\href@noop {} {\bibfield  {journal} {\bibinfo  {journal} {SIAM Journal on
				Applied Dynamical Systems}\ }\textbf {\bibinfo {volume} {4}},\ \bibinfo
		{pages} {78} (\bibinfo {year} {2005})}\BibitemShut {NoStop}%
	\bibitem [{\citenamefont {Nakao}(2016)}]{nakao2016phase}%
	\BibitemOpen
	\bibfield  {author} {\bibinfo {author} {\bibfnamefont {H.}~\bibnamefont
			{Nakao}},\ }\href@noop {} {\bibfield  {journal} {\bibinfo  {journal}
			{Contemp. Phys.}\ }\textbf {\bibinfo {volume} {57}},\ \bibinfo {pages} {188}
		(\bibinfo {year} {2016})}\BibitemShut {NoStop}%
	\bibitem [{\citenamefont {Pietras}\ and\ \citenamefont
		{Daffertshofer}(2019)}]{pietras2019network}%
	\BibitemOpen
	\bibfield  {author} {\bibinfo {author} {\bibfnamefont {B.}~\bibnamefont
			{Pietras}}\ and\ \bibinfo {author} {\bibfnamefont {A.}~\bibnamefont
			{Daffertshofer}},\ }\href@noop {} {\bibfield  {journal} {\bibinfo  {journal}
			{Phys. Rep.}\ } (\bibinfo {year} {2019})}\BibitemShut {NoStop}%
	\bibitem [{\citenamefont {Matheny}\ \emph {et~al.}(2019)\citenamefont
		{Matheny}, \citenamefont {Emenheiser}, \citenamefont {Fon}, \citenamefont
		{Chapman}, \citenamefont {Salova}, \citenamefont {Rohden}, \citenamefont
		{Li}, \citenamefont {{de Badyn}}, \citenamefont {P{\'o}sfai}, \citenamefont
		{{Duenas-Osorio}} \emph {et~al.}}]{matheny2019exotic}%
	\BibitemOpen
	\bibfield  {author} {\bibinfo {author} {\bibfnamefont {M.~H.}\ \bibnamefont
			{Matheny}}, \bibinfo {author} {\bibfnamefont {J.}~\bibnamefont {Emenheiser}},
		\bibinfo {author} {\bibfnamefont {W.}~\bibnamefont {Fon}}, \bibinfo {author}
		{\bibfnamefont {A.}~\bibnamefont {Chapman}}, \bibinfo {author} {\bibfnamefont
			{A.}~\bibnamefont {Salova}}, \bibinfo {author} {\bibfnamefont
			{M.}~\bibnamefont {Rohden}}, \bibinfo {author} {\bibfnamefont
			{J.}~\bibnamefont {Li}}, \bibinfo {author} {\bibfnamefont {M.~H.}\
			\bibnamefont {{de Badyn}}}, \bibinfo {author} {\bibfnamefont
			{M.}~\bibnamefont {P{\'o}sfai}}, \bibinfo {author} {\bibfnamefont
			{L.}~\bibnamefont {{Duenas-Osorio}}},  \emph {et~al.},\ }\href@noop {}
	{\bibfield  {journal} {\bibinfo  {journal} {Science}\ }\textbf {\bibinfo
			{volume} {363}},\ \bibinfo {pages} {eaav7932} (\bibinfo {year}
		{2019})}\BibitemShut {NoStop}%
	\bibitem [{\citenamefont {Komarov}\ and\ \citenamefont
		{Pikovsky}(2013)}]{komarov2013dynamics}%
	\BibitemOpen
	\bibfield  {author} {\bibinfo {author} {\bibfnamefont {M.}~\bibnamefont
			{Komarov}}\ and\ \bibinfo {author} {\bibfnamefont {A.}~\bibnamefont
			{Pikovsky}},\ }\href@noop {} {\bibfield  {journal} {\bibinfo  {journal}
			{Phys. Rev. Lett.}\ }\textbf {\bibinfo {volume} {110}},\ \bibinfo {pages}
		{134101} (\bibinfo {year} {2013})}\BibitemShut {NoStop}%
	\bibitem [{\citenamefont {Ashwin}\ and\ \citenamefont
		{Rodrigues}(2016)}]{ashwin2016hopf}%
	\BibitemOpen
	\bibfield  {author} {\bibinfo {author} {\bibfnamefont {P.}~\bibnamefont
			{Ashwin}}\ and\ \bibinfo {author} {\bibfnamefont {A.}~\bibnamefont
			{Rodrigues}},\ }\href@noop {} {\bibfield  {journal} {\bibinfo  {journal}
			{Phys. D}\ }\textbf {\bibinfo {volume} {325}},\ \bibinfo {pages} {14}
		(\bibinfo {year} {2016})}\BibitemShut {NoStop}%
	\bibitem [{\citenamefont {Ashwin}\ \emph {et~al.}(2016)\citenamefont {Ashwin},
		\citenamefont {Bick},\ and\ \citenamefont {Burylko}}]{ashwin2016identical}%
	\BibitemOpen
	\bibfield  {author} {\bibinfo {author} {\bibfnamefont {P.}~\bibnamefont
			{Ashwin}}, \bibinfo {author} {\bibfnamefont {C.}~\bibnamefont {Bick}}, \ and\
		\bibinfo {author} {\bibfnamefont {O.}~\bibnamefont {Burylko}},\ }\href@noop
	{} {\bibfield  {journal} {\bibinfo  {journal} {Front Appl. Math. Stat.}\
		}\textbf {\bibinfo {volume} {2}},\ \bibinfo {pages} {7} (\bibinfo {year}
		{2016})}\BibitemShut {NoStop}%
	\bibitem [{\citenamefont {Le{\'o}n}\ and\ \citenamefont
		{Paz{\'o}}(2019)}]{leon2019phase}%
	\BibitemOpen
	\bibfield  {author} {\bibinfo {author} {\bibfnamefont {I.}~\bibnamefont
			{Le{\'o}n}}\ and\ \bibinfo {author} {\bibfnamefont {D.}~\bibnamefont
			{Paz{\'o}}},\ }\href@noop {} {\bibfield  {journal} {\bibinfo  {journal}
			{Phys. Rev. E}\ }\textbf {\bibinfo {volume} {100}},\ \bibinfo {pages}
		{012211} (\bibinfo {year} {2019})}\BibitemShut {NoStop}%
	\bibitem [{\citenamefont {Bick}\ \emph {et~al.}(2016)\citenamefont {Bick},
		\citenamefont {Ashwin},\ and\ \citenamefont {Rodrigues}}]{bick2016chaos}%
	\BibitemOpen
	\bibfield  {author} {\bibinfo {author} {\bibfnamefont {C.}~\bibnamefont
			{Bick}}, \bibinfo {author} {\bibfnamefont {P.}~\bibnamefont {Ashwin}}, \ and\
		\bibinfo {author} {\bibfnamefont {A.}~\bibnamefont {Rodrigues}},\ }\href@noop
	{} {\bibfield  {journal} {\bibinfo  {journal} {Chaos}\ }\textbf {\bibinfo
			{volume} {26}},\ \bibinfo {pages} {094814} (\bibinfo {year}
		{2016})}\BibitemShut {NoStop}%
	\bibitem [{\citenamefont {Bick}(2018)}]{bick2018heteroclinic}%
	\BibitemOpen
	\bibfield  {author} {\bibinfo {author} {\bibfnamefont {C.}~\bibnamefont
			{Bick}},\ }\href@noop {} {\bibfield  {journal} {\bibinfo  {journal} {Phys.
				Rev. E}\ }\textbf {\bibinfo {volume} {97}},\ \bibinfo {pages} {050201}
		(\bibinfo {year} {2018})}\BibitemShut {NoStop}%
	\bibitem [{\citenamefont {Bick}(2019)}]{bick2019heteroclinic}%
	\BibitemOpen
	\bibfield  {author} {\bibinfo {author} {\bibfnamefont {C.}~\bibnamefont
			{Bick}},\ }\href@noop {} {\bibfield  {journal} {\bibinfo  {journal} {J Nonlin
				Sci}\ } (\bibinfo {year} {2019})}\BibitemShut {NoStop}%
	\bibitem [{\citenamefont {Kirkpatrick}\ and\ \citenamefont
		{Thirumalai}(1987)}]{kirkpatrick1987dynamics}%
	\BibitemOpen
	\bibfield  {author} {\bibinfo {author} {\bibfnamefont {T.~R.}\ \bibnamefont
			{Kirkpatrick}}\ and\ \bibinfo {author} {\bibfnamefont {D.}~\bibnamefont
			{Thirumalai}},\ }\href@noop {} {\bibfield  {journal} {\bibinfo  {journal}
			{Phys. Rev. Lett.}\ }\textbf {\bibinfo {volume} {58}},\ \bibinfo {pages}
		{2091} (\bibinfo {year} {1987})}\BibitemShut {NoStop}%
	\bibitem [{\citenamefont {De~Domenico}\ \emph {et~al.}(2016)\citenamefont
		{De~Domenico}, \citenamefont {Granell}, \citenamefont {Porter},\ and\
		\citenamefont {Arenas}}]{de2016physics}%
	\BibitemOpen
	\bibfield  {author} {\bibinfo {author} {\bibfnamefont {M.}~\bibnamefont
			{De~Domenico}}, \bibinfo {author} {\bibfnamefont {C.}~\bibnamefont
			{Granell}}, \bibinfo {author} {\bibfnamefont {M.~A.}\ \bibnamefont {Porter}},
		\ and\ \bibinfo {author} {\bibfnamefont {A.}~\bibnamefont {Arenas}},\
	}\href@noop {} {\bibfield  {journal} {\bibinfo  {journal} {Nat. Phys.}\
		}\textbf {\bibinfo {volume} {12}},\ \bibinfo {pages} {901} (\bibinfo {year}
		{2016})}\BibitemShut {NoStop}%
	\bibitem [{\citenamefont {Lambiotte}\ \emph {et~al.}(2019)\citenamefont
		{Lambiotte}, \citenamefont {Rosvall},\ and\ \citenamefont
		{Scholtes}}]{lambiotte2019networks}%
	\BibitemOpen
	\bibfield  {author} {\bibinfo {author} {\bibfnamefont {R.}~\bibnamefont
			{Lambiotte}}, \bibinfo {author} {\bibfnamefont {M.}~\bibnamefont {Rosvall}},
		\ and\ \bibinfo {author} {\bibfnamefont {I.}~\bibnamefont {Scholtes}},\
	}\href@noop {} {\bibfield  {journal} {\bibinfo  {journal} {Nat. Phys.}\ ,\
			\bibinfo {pages} {1}} (\bibinfo {year} {2019})}\BibitemShut {NoStop}%
\end{thebibliography}
\end{document}